\documentclass[aps,prd,reprint,twocolumn,superscriptaddress, floatfix,preprintnumbers]{revtex4-2}
\usepackage{graphicx,amsmath,amsfonts,amssymb,amsthm,xr, physics}
\usepackage{color,dsfont,upgreek}
\usepackage{mathrsfs}
\usepackage{siunitx}
\usepackage{mathtools}
\usepackage{bbold}
\usepackage{slashed}
\usepackage{float}
\usepackage[caption=false]{subfig}
\usepackage[dvipsnames]{xcolor}
\usepackage[bookmarks=true,colorlinks,linkcolor=OrangeRed,urlcolor=NavyBlue,citecolor=RoyalBlue]{hyperref}
\usepackage[capitalize]{cleveref}
\usepackage{orcidlink}

\usepackage{dcolumn}
\usepackage{bm}

\usepackage{lineno}

\newcommand{\mitus}{Center for Theoretical Physics - a Leinweber Institute, Massachusetts Institute of Technology, Cambridge, MA 02139, USA}

\begin{document}

\title{Neural quantum states for non-Abelian lattice gauge theories with dynamical fermions}

\author{Gabriel Rouxinol$^{\orcidlink{0009-0004-8147-9814}}$}
\affiliation{Department of Physics and Arnold Sommerfeld Center for Theoretical Physics (ASC), Ludwig Maximilian University of Munich, 80333 Munich, Germany}
\affiliation{Munich Center for Quantum Science and Technology (MCQST), 80799 Munich, Germany}

\author{Julian Bender${}^{\orcidlink{0000-0003-4920-7849}}$}
\affiliation{\mitus}

\author{Michele Grossi$^{\orcidlink{0000-0003-1718-1314}}$}
\affiliation{European Organisation for Nuclear Research (CERN), 1211 Geneva, Switzerland}

\author{Patrick Emonts$^{\orcidlink{0000-0002-7274-4071}}$}
\affiliation{Institute for Complex Quantum Systems, Ulm University, 89069 Ulm, Germany}
\affiliation{Center for Integrated Quantum Science and Technology (IQST), Ulm-Stuttgart, Germany}

\author{Jad C.~Halimeh${}^{\orcidlink{0000-0002-0659-7990}}$}
\email{jad.halimeh@lmu.de}
\affiliation{Department of Physics and Arnold Sommerfeld Center for Theoretical Physics (ASC), Ludwig Maximilian University of Munich, 80333 Munich, Germany}
\affiliation{Max Planck Institute of Quantum Optics, 85748 Garching, Germany}
\affiliation{Munich Center for Quantum Science and Technology (MCQST), 80799 Munich, Germany}
\affiliation{Department of Physics, College of Science and Technology, Kyung Hee University, Seoul 02447, Republic of Korea}

\date{\today}

\begin{abstract}
Determining the ground state of non-Abelian lattice gauge theories coupled to dynamical fermions is key to understanding confinement and the phase structure of gauge--matter systems. We present a variational Monte Carlo framework for the ground state of the untruncated fully-continuous SU$(2)$ lattice gauge theory coupled to dynamical staggered fermions on an $L\times L$ square lattice. We work in the magnetic basis with a neural-network representation of the gauge wavefunction. The fermions are described by a gauge-covariant Gaussian fermionic correction built on a fixed N\'eel reference state where, for each sampled gauge configuration $\mathbf{U}$, the correction is generated by a Hermitian operator. This operator is constructed from short Wilson lines and the eigenvectors of the mass--hopping Hamiltonian, with number of variational parameters polynomial in the system size. This Gaussian structure also gives analytical expressions for all fermionic contributions to the energy and related observables in terms of the fermion occupation matrix. The results are validated against strong-coupling perturbation theory, where they recover the expected effective antiferromagnetic spin Hamiltonian. Using this framework, we map a coarse ground state phase diagram in the plane of independent electric and magnetic couplings $(g^2, \lambda)$ and show that a hysteresis analysis can identify the existence of phase transitions. Restoring the physical relation $\lambda=4/g^2$, we characterize how increasing the system size and changing the electric coupling $g^2$ move the state away from the reference N\'eel state, for lattice sizes $L=4,6,8$. More broadly, the method offers a sign-problem-free variational framework for continuous non-Abelian gauge groups with dynamical matter that should extend to other matter content and higher-dimensional lattices.
\end{abstract}

\maketitle

\section{Introduction}
\label{sec:intro}

The Standard Model of Particle Physics (SM) is currently our most complete description of three of the four fundamental forces in Nature~\cite{Weinberg1995QuantumTheoryFields, Weinberg:2004kv}. 
Its formulation utilizes the language of gauge theories, where matter and gauge fields are introduced in a way invariant to gauge transformations~\cite{peskin2018introduction, schwartz2014quantum, srednicki2007quantum}. 
Although this formulation allows for high-precision predictions using perturbative theory at high energies, the strongly coupled low-energy physics can only be described accurately in a discretized, non-perturbative way~\cite{grossAsymptoticallyFreeGauge1973,wilsonConfinementQuarks1974}. 
A distinct approach to describe the theory in these regimes was proposed by Kogut and Susskind~\cite{kogutHamiltonianFormulationWilsons1975}. 
This Hamiltonian formulation works in continuous real time, in contrast to the Lagrangian one in Euclidean space with discrete imaginary time. 
Lattice Gauge Theories (LGTs) also arise independently in the study of condensed matter problems, where they are effective descriptions of strongly correlated systems and quantum spin liquids~\cite{baskaranResonatingValenceBond1987, affleckLargenLimitHeisenbergHubbard1988,wenMeanfieldTheorySpinliquid1991, leeDopingMottInsulator2006}. 

Among numerical approaches to LGTs, Monte Carlo~\cite{creutzMonteCarloStudy1979} in the Euclidean formulation has produced important results, including quark masses and the QCD phase diagram~\cite{wilsonConfinementQuarks1974, carlsonQuantumMonteCarlo2015, alltonQuarkMassesLattice1994, fodorNewMethodStudy2002}. 
Nevertheless, it has its limitations due to the infamous sign-problem~\cite{troyerComputationalComplexityFundamental2005}, which occurs in the presence of a topological $\theta$-term~\cite{vicariThthDependenceSUNSUN2009} or of dynamical fermions with a finite chemical potential~\cite{philipsenLatticeQCDFinite2007}, blocking access to large regions of physically relevant parameter space. 
Solutions based on Machine Learning (ML)~\cite{boydaApplicationsMachineLearning2022, alexandruComplexPathsSign2022} methods have been proposed to tackle this problem, by introducing flow-based techniques to sampling~\cite{kanwarEquivariantFlowBasedSampling2020, albergoFlowbasedSamplingFermionic2021, lawrenceNormalizingFlowsRealtime2021} or by using machine learning to deform the integration contour or optimize the action parameters directly~\cite{shanahanMachineLearningAction2018, hisayoshiPathOptimizationMethod2025}. The Hamiltonian formulation instead provides a natural framework for methods that are sign-problem-free by construction.

In the Hamiltonian formulation, quantum simulation has demonstrated the ability to probe non-equilibrium dynamics, confinement, and thermalization in Abelian and non-Abelian gauge groups~\cite{dimeglioQuantumComputingHighEnergy2024, bauerQuantumSimulationHighEnergy2023, byrnesSimulatingLatticeGauge2006, halimehQuantumSimulationOutofequilibrium2025,dalmonteLatticeGaugeTheory2016, aidelsburgerColdAtomsMeet2021a, zoharQuantumSimulationsLattice2015, barataMediumInducedJet2022, barataQuantumSimulationInmedium2023, barataRealtimeDynamicsHyperon2024, halimehAchievingQuantumField2022,halimehTuningTopological$ensuremaththeta$Angle2022, klcoQuantumclassicalComputationSchwinger2018, zhangObservationMicroscopicConfinement2025, farrellScalableCircuitsPreparing2024, farrellPreparationsQuantumSimulations2023, atasSU2HadronsQuantum2021, thanPhaseDiagramQuantum2025, angelidesFirstorderPhaseTransition2025, schuhmacherObservationHadronScattering2025, davoudiQuantumComputationHadron2025, liFrameworkQuantumSimulations2026, suraceLatticeGaugeTheories2020, desaulesProminentQuantumManybody2023, schweizerFloquetApproachZ22019, suObservationManybodyScarring2023, zhouThermalizationDynamicsGauge2022, ciavarellaTrailheadQuantumSimulation2021, ciavarellaQuantumSimulationLattice2023}. 
Recently, both digital and analog implementations have extended these results to $2+1$ and $3+1$D~\cite{banerjeeAtomicQuantumSimulation2013, tagliacozzoSimulationNonAbelianGauge2013, kan3+1DThetaTermLattice2022, magnificoLatticeQuantumElectrodynamics2021, zoharQuantumSimulationLattice2021, leeQuantumComputingEnergy2025, turroClassicalQuantumComputing2024, gyawaliObservationDisorderfreeLocalization2025, mildenbergerConfinement$$mathbbZ_2$$Lattice2025, gonzalez-cuadraObservationStringBreaking2025, crippaAnalysisConfinementString2024, ciavarellaTruncationUncertaintiesAccurate2026, ebnerEntanglementEntropy$2+1$dimensional2024, pacianiQuantumSimulationFermionic2025, joshiObservationGenuine$2+1$D2026, xuObservationGlueballExcitations2026}. 
However, they require a truncation of the gauge group, which limits their ability to probe the continuum limit, and remain bounded by current hardware capabilities~\cite{banulsSimulatingLatticeGauge2020, klcoStandardModelPhysics2022}. 
Tensor networks have emerged as a classical numerical approach to the Hamiltonian formulation, as they can access real time evolution and encode the quantum state efficiently through the entanglement structure~\cite{whiteDensityMatrixFormulation1992,whiteDensitymatrixAlgorithmsQuantum1993, schollwockDensitymatrixRenormalizationGroup2011, orusPracticalIntroductionTensor2014,montangeroIntroductionTensorNetwork2018}. 
They have achieved excellent results in $1+1$D problems~\cite{banulsMassSpectrumSchwinger2013, banulsEfficientBasisFormulation2017, belyanskyHighEnergyCollisionQuarks2024, rouxinolSchwingerModelDynamical2026, ricoTensorNetworksLattice2014, rigobelloEntanglementGeneration$1+1mathrmD$2021, salaVariationalStudyU12018, silviTensorNetworkSimulation2019}, and even in some higher dimensional problems, although their computational cost scales unfavorably with entanglement in $2+1$ and $3+1$D systems~\cite{cataldiSimulating$2+1mathrmD$SU22024, cataldiHamiltonianLatticeGauge2025, cobosRealTimeDynamics2025, emontsFindingGroundState2023, felserTwoDimensionalQuantumLinkLattice2020, krinitsinTimeEvolutionQuantum2025, magnificoTensorNetworksLattice2025, orusTensorNetworksComplex2019, osborneDisorderFreeLocalization$2+1$D2023, osborneQuantumManyBodyScarring2024, xuStringBreakingDynamics2025, cataldiRealTimeStringDynamics2025}.

A further strategy is Variational Monte Carlo (VMC)~\cite{carleoSolvingQuantumManybody2017, mcmillanGroundStateLiquid1965, arisueVariationalStudyVacuum1983, chinExactGroundstateProperties1985}, where a variational ansatz is optimized to minimize the energy expectation value. 
This technique, together with the development of neural network tools for quantum many-body problems has resulted in improvements over previous numerical tools for many fields~\cite{medvidovicNeuralnetworkQuantumStates2024, pfauAccurateComputationQuantum2024, wuRealNeuralNetwork2023, wuNNQSTransformerEfficientScalable2023, liImprovedOptimizationNeuralnetwork2024, wangVariationalOptimizationAmplitude2024, nysRealtimeQuantumDynamics2024, louNeuralWaveFunctions2024, luVariationalNeuralTensor2025, wuDeepQuarkDeepNeuralNetworkApproach2026}. 
Still, their usage for LGTs has been limited~\cite{bodendorferVariationalMonteCarlo2025, favoniApplicationsLatticeGauge2022, luoGaugeEquivariantNeural2021, luoGaugeinvariantAnyonicsymmetricAutoregressive2023, rayatGraphNeuralNetworks2026}, as building a gauge-invariant variational state which is physically expressive enough to describe the ground state of the theory is a hard task. 
Recently, neural-network approaches have been presented for a pure SU$(2)$ gauge theory in $2+1$ and $3+1$D~\cite{spriggsAccurateGroundStates2026}, by using ML to improve on purely variational results. 
Additionally, there are approaches for simpler theories, such as U$(1)$ or $\mathbb{Z}_2$ with dynamical matter, mainly utilizing Gaussian fermionic states~\cite{benderVariationalMonteCarlo2023, emontsFindingGroundState2023}. 
In this work, we combine the two approaches by merging the pure-gauge SU$(2)$ VMC and the U$(1)$ VMC with dynamical fermions to develop an algorithm for the SU$(2)$ LGT with matter. 
This allows us to study an important model for both high-energy and condensed matter physics, as it is the simplest non-Abelian model in $2+1$D exhibiting confinement and a rich phase structure, serving as a stepping stone toward QCD~\cite{tepermathrmSUNGaugeTheories1998,karabaliGaugeinvariantHamiltonianAnalysis1996,feynmanQualitativeBehaviorYangMills1981}.

We start this work by presenting the Hamiltonian of the SU$(2)$ LGT in $2+1$D in Sec.~\ref{sec:model}. 
After that, we present the ansatz we use, its motivation and main properties in Sec.~\ref{sec:ansatz_description}. 
This is followed by the derivation of the analytical expression of the expectation values used in our work in Sec.~\ref{sec:analytical_expressions}.
Following that, the training parameters and protocols used in our work are extensively described in Sec.~\ref{sec:implementation_details}. 
We then study the model in the strong-coupling limit in Sec.~\ref{sec:high_coupling_limit}. 
Secs.~\ref{sec:phase_diagram} and~\ref{sec:physical_line} present the numerical results, with emphasis on the variational and finite-size diagnostics, while their physical interpretation is developed in the companion Letter~\cite{rouxinolMachineLearningBased2026}.
We finish by summarizing our findings and presenting future directions in Sec.~\ref{sec:discussion}.

\section{Model and Hamiltonian}
\label{sec:model}
We study a $2+1$-dimensional SU$(2)$ LGT coupled to dynamical fermions on an $L\times L$ square lattice, with $N=L^2$ sites and periodic boundary conditions.
We will work with the Hamiltonian formulation of this theory using staggered fermions~\cite{kogutHamiltonianFormulationWilsons1975}:

\begin{align} \label{eqn:HamiltonianLGT}
	\hat{H} &= \hat{H}_E+ \hat{H}_B + \hat{H}_m + \hat{H}_t,\\\nonumber
    \hat{H}_E &= \frac{g^2}{2a_s} \sum_{\mathbf{n},k,a} \hat{E}_{\mathbf{n},\boldsymbol{\mu}_k}^a \hat{E}_{\mathbf{n},\boldsymbol{\mu}_k}^a,\\\nonumber
    \hat{H}_B&= \frac{\lambda}{a_s}\sum_{\mathbf{n}} \left(1 - \frac{1}{2} \Tr \hat{P}_{\mathbf{n}, \Box}\right),\\\nonumber
	\hat{H}_m&= m\sum_{\mathbf{n}, \alpha}(-1)^{n_x+n_y}\hat{\psi}_{\mathbf{n}, \alpha}^\dagger\hat{\psi}_{\mathbf{n}, \alpha},\\\nonumber
	\hat{H}_t&= -\frac{it}{2a_s} \sum_{\mathbf{n},k,\alpha, \beta}  \left(\hat{\psi}_{\mathbf{n}, \alpha}^\dagger \hat{U}^{\alpha\beta}_{\mathbf{n},\boldsymbol{\mu}_k} \hat{\psi}_{\mathbf{n}+\boldsymbol{\mu}_k, \beta} \eta_{\mathbf{n},\boldsymbol{\mu}_k} - \text{H.c.}\right),
\end{align}
where $\mathbf{n}=(n_x,n_y)$ labels the vertices on the lattice, $\boldsymbol{\mu}_k$ the unit vector in direction $k \in (x,y)$. 
The link connecting $\mathbf{n}$ with $\mathbf{n}+\boldsymbol{\mu}_k$ is denoted by $(\mathbf{n}, \boldsymbol{\mu}_k)$, and the gauge operator on that link is $\hat{U}_{\mathbf{n},\boldsymbol{\mu}_k}$. 
The factor $\eta_{\mathbf{n},\boldsymbol{\mu}_k}$ is the staggering factor defined as $\eta_{\mathbf{n},\boldsymbol{\mu}_y} =(-1)^{n_x}$ on vertical links and $\eta_{\mathbf{n},\boldsymbol{\mu}_x}=1$ otherwise. 
The fermion field is represented in the staggered formulation by the annihilation operator $\psi_{\mathbf{n},\alpha}$, which follows the standard relations $\{\psi_{\mathbf{n},\alpha},{\psi^{\dagger}_{\mathbf{n}^{\prime},\beta}}\}\!=\! 
\delta_{\mathbf{n},\mathbf{n}^{\prime}} \delta_{\alpha, \beta}$ and $\{\psi_{\mathbf{n},\alpha},{\psi_{\mathbf{n}^{\prime},\beta}}\}\!=\!\{\psi^\dagger_{\mathbf{n},\alpha},{\psi^\dagger_{\mathbf{n}^{\prime},\beta}}\}\!=\!0$, with $\alpha$ and $\beta$ indices in the fundamental irreducible representation of SU$(2)$.
The set of all lattice links will be denoted by $\mathbf{U}$.
We also define $g$ as the coupling constant, $a_s$ as the lattice spacing, which we set to unity, $t$ as the hopping parameter and $m$ as the fermion mass. We will take $t=1.0$ and $m=0.5$, except when stated otherwise. 
In the standard lattice-coupling convention, the magnetic and electric pre-factors are related by $\lambda=\frac{4}{g^2}$. 
However, we will take them as independent parameters. 
Finally, $\hat{P}_{\mathbf{n}, \Box} = \sum_{\alpha,\beta,\gamma,\delta}\hat{U}^{\alpha\beta}_{\mathbf{n},\boldsymbol{\mu}_x}\hat{U}^{\beta\gamma}_{\mathbf{n}+\boldsymbol{\mu}_x,\boldsymbol{\mu}_y}\hat{U}^{\gamma\delta\dagger}_{\mathbf{n}+\boldsymbol{\mu}_y,\boldsymbol{\mu}_x}\hat{U}^{\delta\alpha\dagger}_{\mathbf{n},\boldsymbol{\mu}_y}$ is the plaquette term and $ \hat{E}_{\mathbf{n},\boldsymbol{\mu}_k}^a$ is either the left or right generator of SU$(2)$, such that they respectively verify either $[\hat{E}_{\mathbf{n},\boldsymbol{\mu}_k,L}^a, \hat{U}_{\mathbf{n},\boldsymbol{\mu}_k}]=\frac{1}{2}\sigma^a\hat{U}_{\mathbf{n},\boldsymbol{\mu}_k}$ or $[\hat{E}_{\mathbf{n},\boldsymbol{\mu}_k,R}^a, \hat{U}_{\mathbf{n},\boldsymbol{\mu}_k}]=\hat{U}_{\mathbf{n},\boldsymbol{\mu}_k}\frac{1}{2}\sigma^a$, with $\sigma^a$ the Pauli matrices, and $a$ an index of SU$(2)$ in the adjoint representation, while satisfying the constraint $\sum_a \hat{E}_{\mathbf{n},\boldsymbol{\mu}_k,L}^a\hat{E}_{\mathbf{n},\boldsymbol{\mu}_k,L}^a=\sum_a\hat{E}_{\mathbf{n},\boldsymbol{\mu}_k,R}^a\hat{E}_{\mathbf{n},\boldsymbol{\mu}_k,R}^a$. 
As the Hamiltonian is invariant under the gauge transformation $\hat{U}_{\mathbf{n},\boldsymbol{\mu}_k} \to \Omega_\mathbf{n} \hat{U}_{\mathbf{n},\boldsymbol{\mu}_k} \Omega_{\mathbf{n}+\boldsymbol{\mu}_k}^\dagger$, $\hat{\psi}_{\mathbf{n}, \alpha} \to \Omega^{\alpha, \beta}_\mathbf{n} \hat{\psi}_{\mathbf{n}, \beta}$, any wavefunction representing a physical state must be invariant under the same transformation, with no possibility of spontaneously breaking the gauge symmetry~\cite{elitzurImpossibilitySpontaneouslyBreaking1975}. 
Any physical, gauge-invariant state $\ket{\text{phys}}$ satisfies the lattice version of Gauss's law $\hat{G}^a_\mathbf{n}\ket{\text{phys}}=0$, where $\hat{G}^a_\mathbf{n} = \sum_k\left[\hat{E}_{\mathbf{n},\boldsymbol{\mu}_k,L}^a - \hat{E}^a_{\mathbf{n}-\boldsymbol{\mu}_k,\boldsymbol{\mu}_k,R}\right] - \hat{\rho}^a_\mathbf{n}$ and $\hat{\rho}^a_\mathbf{n} = \sum_{\alpha,\beta}\hat{\psi}_{\mathbf{n}, \alpha}^\dagger \frac{(\sigma^a)^{\alpha,\beta}}{2} \hat{\psi}_{\mathbf{n}, \beta}$. 
We also define the plaquette average as
\begin{equation}
\langle\cos\hat B_p\rangle=\frac{1}{2N}\Big\langle\sum_{\mathbf n}\Tr \hat P_{\mathbf n,\Box}\Big\rangle,
\label{eqn:cosDefinition}
\end{equation}
which will be the observable used to characterize the magnetic flux phases, as it will take the values $\langle\cos\hat B_p\rangle=\pm 1$ for a maximally ordered magnetic phase and $\langle\cos\hat B_p\rangle=0$ for a completely disordered magnetic phase.

\section{Variational ansatz and its gauge covariance}
\label{sec:ansatz_description}
Representing general, gauge-invariant states while dealing with an infinite-dimensional basis is an open problem in computational methods for LGTs. 
This is addressed, in contemporary methods, via truncations or by building gauge-invariant bases~\cite{dalmonteLatticeGaugeTheory2016, zoharQuantumSimulationLattice2021, halimehQuantumSimulationOutofequilibrium2025, magnificoTensorNetworksLattice2025, ciavarellaTruncationUncertaintiesAccurate2026}. 
The latter becomes intractable in higher dimensions, while the former reveals itself to be constraining when going to the continuum limit. 
A solution for pure SU$(2)$ has been proposed in Ref.~\cite{spriggsAccurateGroundStates2026}, where the gauge links are directly parametrized, allowing a study of the full gauge group in the so-called magnetic basis. 
We expand on this work by including dynamical fermions. 
Its basic building block is a spherical representation of the link variables using parameters $(\rho, \theta, \phi)$, as
\begin{equation}
    	\hat{U}_{\mathbf{n},\boldsymbol{\mu}_k} = \cos\left(\frac{\rho}{2}\right)\mathbb{I} - i\Vec{n}\cdot \Vec{\sigma} \sin \left(\frac{\rho}{2}\right),
        \label{eqn:UdefinitionS3}
\end{equation}
where $\Vec{n}=(\sin(\theta)\cos(\phi), \sin(\theta)\sin(\phi), \cos(\theta))$. 
In this representation, the electric field operator becomes the Laplace-Beltrami operator on the SU$(2)$ group manifold $S^3$, meaning $\hat{E}^2=-\frac{1}{4}\nabla^2_{S^3}$, at each link. 
The gauge wavefunction $\Psi_G(\mathbf{U})$ is a combination of a two-body Jastrow ansatz with a convolutional neural network. 
The two-body term of the Jastrow ansatz is built in a translationally invariant manner, allowing smaller system sizes to be loaded into larger system sizes. 
The convolutional neural network allows the wavefunction to capture higher correlations beyond those captured by the Jastrow ansatz. 
More details on the ansatz used and the neural network trained to minimize the energy are explained in the original work~\cite{spriggsAccurateGroundStates2026}. 
To extend this model to one that considers matter, we use an approach that has been demonstrated for a U$(1)$ LGT in $2+1$D~\cite{benderVariationalMonteCarlo2023}, where the total state is defined as $\ket{\Psi}=\int \mathcal{D}\mathbf{U}\Psi_G(\mathbf{U})\ket{\Psi_F(\mathbf{U})}\ket{\mathbf{U}},$ with $\mathcal{D}\mathbf{U}=\prod_{\mathbf{n},k}dU_{\mathbf{n},\boldsymbol{\mu}_k}$ and $\ket{\Psi_F(\mathbf{U})}$ a fermion state specified by the gauge configuration $\mathbf{U}$ and built in a gauge-covariant way. 
In this structure, the state $\ket{\Psi_F(\mathbf{U})}$ determines the low-energy physics of the fermionic Hamiltonian $\hat{H}_\text{fer}\equiv  \hat{H}_E+ \hat{H}_m + \hat{H}_t \equiv\hat{H}_E+\hat{H}_{\mathrm{MH}}$. 
The electric part of the Hamiltonian needs to be considered in the fermionic Hamiltonian, as the Laplacian that defines it also acts on the matter state. 
We will later split the expectation value of $\hat H_E$ into an only gauge wavefunction dependent part $\langle\hat H_E\rangle_{GG}$, a pure fermionic state dependent part $\langle\hat H_E\rangle_{\mathrm{FF}}$, and one that mixes both $\langle\hat H_E\rangle_{\mathrm{GF}}$. 
However, the fermionic state depends on the sampled gauge configurations, so the full $\langle \hat H_E\rangle$ is considered in the gauge training, while only $\langle\hat H_E\rangle_{\mathrm{FF}}$ and $\langle\hat H_E\rangle_{\mathrm{GF}}$ are included in the matter training.  
We take $\ket{\Psi_F(\mathbf{U})}$ to be Gaussian for fixed $\mathbf{U}$, which keeps the number of variational parameters polynomial in the system size. 
Note, however, that $\ket{\Psi_F(\mathbf{U})}$ being Gaussian at every $\mathbf{U}$ does not result in a Gaussian matter state after sampling. 
Once the gauge configurations $\mathbf{U}$ are traced out, the state is a linear combination of Gaussian states rather than a single Gaussian state, which is precisely what allows non-Gaussian properties to be reached~\cite{bravyiComplexityQuantumImpurity2017}.
In the half-filled calculations reported here, the reference state is the gauge-invariant N\'eel state $\ket{\Psi_N}$ with fermion occupation matrix $P_N=[P_N]_{\mathbf{n}',\beta,\mathbf{n},\alpha}= \bra{\Psi_N}\hat{\psi}^\dagger_{\mathbf{n},\alpha}\hat{\psi}_{\mathbf{n}',\beta}\ket{\Psi_N}=\frac{1}{2}\big(1-(-1)^{n_x+n_y}\big)\delta_{\mathbf{n},\mathbf{n}'}\delta_{\alpha,\beta}$. 
In that case, we take $\ket{\Psi_F(\mathbf{U})}=\hat{U}_{\mathrm{corr}}(\mathbf{U})\ket{\Psi_N}$, where $\hat{U}_{\mathrm{corr}}(\mathbf{U})$ is a unitary determined by the gauge configuration $\mathbf{U}$. 
This transformation has to be unitary and gauge covariant. 
For that, we define it as
\begin{equation}
\hat{U}_{\mathrm{corr}}(\mathbf{U})=\exp\bigg(i\sum_{\mathbf{n},\alpha,\mathbf{n}',\beta}\hat{\psi}_{\mathbf{n}, \alpha}^\dagger [H_{\mathrm{full}}(\mathbf{U})]_{\mathbf{n},\mathbf{n}'}^{\alpha\beta} \hat{\psi}_{\mathbf{n}', \beta}\bigg),
\label{eqn:UGSdefinition}
\end{equation}
where the explicit construction of the generator $H_{\mathrm{full}}(\mathbf{U})$ is given in Sec.~\ref{sec:ansatz_construction}.  
If $\ket{\Psi_N}$ is gauge invariant and $H_{\mathrm{full}}(\mathbf{U})$ transforms covariantly under local gauge transformations, then the corresponding many-body operator $\hat{U}_{\mathrm{corr}}(\mathbf{U})$ is built in a gauge-covariant way, and therefore so is $\ket{\Psi_F(\mathbf{U})}$. 
This defines the fermion occupation matrix, with elements $P(\mathbf{U})_{\mathbf{n}',\beta,\mathbf{n},\alpha}= \bra{\Psi_F(\mathbf{U})}\hat{\psi}^\dagger_{\mathbf{n},\alpha}\hat{\psi}_{\mathbf{n}',\beta}\ket{\Psi_F(\mathbf{U})}$, obtained as $P(\mathbf{U})=U_{\mathrm{corr}}(\mathbf{U}) P_N U_{\mathrm{corr}}^\dagger(\mathbf{U})$, with $U_{\mathrm{corr}}(\mathbf{U})=e^{iH_{\mathrm{full}}(\mathbf{U})}$. For $H_{\mathrm{full}}(\mathbf{U})=0$, it reduces to the occupation matrix of the reference state $P_N$.
The generator $H_{\mathrm{full}}(\mathbf{U})$ must then be able to capture the tendency of the hopping Hamiltonian to delocalize the fermions while penalizing the long flux strings that increase the electric-field energy. 
Its explicit construction is given in Sec.~\ref{sec:ansatz_construction}, together with a verification that the ansatz satisfies Gauss's law in Sec.~\ref{sec:gauge_covariance}. 
One useful property of this Gaussian construction is that all expectation values of the fermionic contributions to the Hamiltonian in Eq.~\eqref{eqn:HamiltonianLGT} can be written analytically in terms of $U_{\mathrm{corr}}(\mathbf{U})$ and $P_N$, as summarized in Sec.~\ref{sec:analytical_expressions}. 

\subsection{Ansatz construction}
\label{sec:ansatz_construction}
In this section, we construct an adequate ansatz for the fermions that takes the N\'eel state as the reference state. 
As the N\'eel state already minimizes the mass term and makes the electric terms that couple to the fermionic state vanish, one needs an ansatz to capture the low-energy properties of the hopping Hamiltonian, while still considering the mass Hamiltonian and the cost of creating electric flux strings.
Firstly, our reference state $\ket{\Psi_N}$ doubly occupies the favored sublattice and leaves the other sublattice empty, so each occupied site is a local color singlet. 
Therefore, $P_N$ is gauge independent and commutes with local gauge rotations. 

For each gauge configuration $\mathbf{U}$, the correction directions are extracted from the matrix representation of the mass--hopping Hamiltonian
\begin{equation}
	h_{\mathrm{MH}}(\mathbf{U})=h_m-\frac{it}{2}\,h_g(\mathbf{U}),
\end{equation}
where $h_m$ is the staggered-mass matrix and $h_g(\mathbf{U})$ is the gauge-covariant hopping matrix. 
In the matrix form, they follow $[h_m]_{\mathbf n\alpha,\mathbf n'\beta}=m\delta_{\mathbf{n}, \mathbf{n}'}\delta_{\alpha,\beta}(-1)^{n_x+n_y}$ and $[h_g(\mathbf U)]_{\mathbf n\alpha,\mathbf n +\boldsymbol{\mu}_k\beta}=\eta_{\mathbf{n},\boldsymbol{\mu}_k}U^{\alpha\beta}_{\mathbf{n},\boldsymbol{\mu}_k}$ and $[h_g(\mathbf U)]_{\mathbf n+\boldsymbol{\mu}_k\alpha,\mathbf n \beta}=-\eta_{\mathbf{n},\boldsymbol{\mu}_k}[U^\dagger_{\mathbf{n},\boldsymbol{\mu}_k}]^{\alpha\beta}$, with $\hat{U}_{\mathbf{n}, \boldsymbol{\mu}_k}^{\alpha\beta}\ket{\mathbf{U}}=U_{\mathbf{n}, \boldsymbol{\mu}_k}^{\alpha\beta}\ket{\mathbf{U}}$. 
Let $V_{\mathrm{occ}}(\mathbf{U})$ contain the $n_{\mathrm{occ}}=N_f$ lowest-energy eigenvectors of $h_{\mathrm{MH}}(\mathbf{U})$, and let $V_{\mathrm{unocc}}(\mathbf{U})$ span the complementary unoccupied subspace of dimension $n_{\mathrm{unocc}}=2L^2-N_f$. 
Although the reference projector is fixed to $P_N$, the pair $(V_{\mathrm{occ}},V_{\mathrm{unocc}})$ defines the occupied--unoccupied space in which the correction acts.

Because the fundamental representation of SU$(2)$ is pseudoreal, $h_{\mathrm{MH}}(\mathbf{U})$ is invariant under the antiunitary symmetry
\begin{equation}
	\mathcal{T}=(S\otimes i\sigma^2)K,
\end{equation}
where $S$ is the $N\times N$ diagonal matrix $[S]_{\mathbf{n},\mathbf{n}}=(-1)^{n_x+n_y}$ and $K$ is complex conjugation. 
The pseudoreality relation $(i\sigma^2)U^*(i\sigma^2)^{-1}=U$ compensates the conjugation in color space, while $S$ compensates the sign flip of the explicit $i$ in $-\tfrac{it}{2}h_g(\mathbf{U})$, which connects only sites of opposite parity. Since $\mathcal{T}^2=-1$, the partner $\mathcal{T}v$ of any eigenvector $v$ is orthogonal to it, while sharing the same eigenvalue, hence all eigenvalues of $h_{\mathrm{MH}}(\mathbf{U})$ occur in twofold-degenerate Kramers pairs. 
For an even number of fermions $N_f$, both the occupied and unoccupied sectors split into two-dimensional Kramers subspaces. 
The individual eigenvectors are then basis dependent inside each pair, so the variational coefficients are organized as scalars on occupied-pair/unoccupied-pair $2\times2$ blocks. 
This removes the arbitrary $U(2)$ basis choice within each Kramers pair.

To couple the fermions to short electric-flux structures, we use three Hermitian Wilson aggregates,
\begin{equation}
	W_{d1}(\mathbf{U}),\qquad W_{d2s}(\mathbf{U}),\qquad W_{d2d}(\mathbf{U}),
\end{equation}
built from distance-1 paths, distance-2 straight paths, and distance-2 diagonal paths, respectively. 
Each aggregate includes the corresponding reversed paths so that the result is Hermitian. 
Explicitly, writing $[W_d]_{\mathbf{n},\mathbf{n}'}$ for the $2\times2$ color block connecting sites $\mathbf{n}$ and $\mathbf{n}'$, the nonvanishing blocks are the ordered parallel transporters
\begin{align}
	[W_{d1}]_{\mathbf{n},\,\mathbf{n}+\boldsymbol{\mu}_k}                      & = U_{\mathbf{n},\boldsymbol{\mu}_k}, \nonumber                                                                                                                                     \\
	[W_{d2s}]_{\mathbf{n},\,\mathbf{n}+2\boldsymbol{\mu}_k}                    & = U_{\mathbf{n},\boldsymbol{\mu}_k}\,U_{\mathbf{n}+\boldsymbol{\mu}_k,\boldsymbol{\mu}_k},                                                                                          \\
	[W_{d2d}]_{\mathbf{n},\,\mathbf{n}+\boldsymbol{\mu}_k+\boldsymbol{\mu}_l}  & = U_{\mathbf{n},\boldsymbol{\mu}_k}\,U_{\mathbf{n}+\boldsymbol{\mu}_k,\boldsymbol{\mu}_l} + U_{\mathbf{n},\boldsymbol{\mu}_l}\,U_{\mathbf{n}+\boldsymbol{\mu}_l,\boldsymbol{\mu}_k}, \nonumber
\end{align}
with $k\in\{x,y\}$ in the first two lines and $k\neq l$ in the third, each accompanied by the Hermitian-conjugate block generated by the reversed path so that $W_d=W_d^\dagger$. 
Thus $W_{d1}$ is the Hermitized nearest-neighbor transporter, $W_{d2s}$ the straight two-link transporter to the site two steps away, and $W_{d2d}$ the symmetrized sum of the two $L$-shaped transporters reaching a diagonal site, which makes it sensitive to the plaquette flux enclosed between the two paths. 
The remaining diagonal neighbors $\mathbf{n}+\boldsymbol{\mu}_k-\boldsymbol{\mu}_l$ are included analogously. 
Restricting the construction to a path length of at most $2$ is the truncation used in this work. 
The analysis on how increasing the truncation length affects the quality of the obtained results is left as future work. 
Although this truncation is limited to distance two Wilson lines, its product with the eigenvectors $V_{\mathrm{occ}}$ and $V_{\mathrm{unocc}}$, which are fully delocalized, will capture non-local properties. 
Furthermore, when we exponentiate the generator $H_{\mathrm{full}}(\mathbf{U})$ we generate products of Wilson lines that will produce longer Wilson lines. 
The only limitation in our approach is that the coefficients of longer Wilson lines are determined by the ones of shorter lines. Nevertheless, the gauge field itself is never truncated, since $\Psi_G(\mathbf{U})$ is a function on the full group manifold and its expansion therefore contains all irreducible representations. The path-length truncation restricts only the fermionic correction.

For each $d\in\{d1,d2s,d2d\}$, we define
\begin{align}
	M_d(\mathbf{U})      & =V_{\mathrm{occ}}^\dagger(\mathbf{U})\,W_d(\mathbf{U})\,V_{\mathrm{unocc}}(\mathbf{U}),   \\
	K_d^{oo}(\mathbf{U}) & =V_{\mathrm{occ}}^\dagger(\mathbf{U})\,W_d(\mathbf{U})\,V_{\mathrm{occ}}(\mathbf{U}),     \\
	K_d^{uu}(\mathbf{U}) & =V_{\mathrm{unocc}}^\dagger(\mathbf{U})\,W_d(\mathbf{U})\,V_{\mathrm{unocc}}(\mathbf{U}).
\end{align}
The matrices $M_d$ give the linear occupied--unoccupied directions, while $K_d^{oo}M_d$ and $M_dK_d^{uu}$ give the quadratic corrections retained in the implementation. 
Physically, the linear term $C_d^{(1)}\odot M_d$ generates a direct Wilson-line-mediated hop between the occupied and unoccupied Kramers subspaces, while the quadratic terms $C_d^{(oo)}\odot(K_d^{oo}M_d)$ and $C_d^{(uu)}\odot(M_dK_d^{uu})$ dress that jump with additional same-sector gauge correlations, letting the correction respond to effectively longer flux paths without extending the path-length truncation itself. 
With elementwise multiplication denoted by $\odot$, the occupied--unoccupied correction matrix is
\begin{equation}
	\begin{aligned}
    	A(\mathbf{U})=\sum_{d\in\{d1,d2s,d2d\}} \Big[&
        C_d^{(1)}\odot M_d(\mathbf{U})\\
        &+C_d^{(oo)}\odot\!(K_d^{oo}(\mathbf{U})M_d(\mathbf{U}))\\
        &+C_d^{(uu)}\odot\!(M_d(\mathbf{U})K_d^{uu}(\mathbf{U}))\Big].
	\end{aligned}
\end{equation}
This gives $9$ channels in total: three linear channels $M_d$, three occupied--occupied quadratic channels $K_d^{oo}M_d$, and three unoccupied--unoccupied quadratic channels $M_dK_d^{uu}$. 
Each coefficient matrix $C_d^{(1)}$, $C_d^{(oo)}$, and $C_d^{(uu)}$ has size $n_{\mathrm{occ}}\times n_{\mathrm{unocc}}$ and is constant on every occupied-pair/unoccupied-pair $2\times2$ block. 
The entry on each $2\times2$ block is a variational parameter optimized during training.

The corresponding Hermitian generator is
\begin{equation}
	H_{\mathrm{full}}(\mathbf{U})
	=V_{\mathrm{occ}}(\mathbf{U})\,A(\mathbf{U})\,V_{\mathrm{unocc}}^\dagger(\mathbf{U}) +\mathrm{H.c.}
\end{equation}
By construction, $H_{\mathrm{full}}(\mathbf{U})$ lies entirely in the off-diagonal block between the occupied and unoccupied subspaces.

The many-body Gaussian correction unitary is
\begin{equation}
	\hat{U}_{\mathrm{corr}}(\mathbf{U})=\exp\bigg(i\sum_{\mathbf{n},\alpha,\mathbf{n}',\beta}\hat{\psi}_{\mathbf{n}, \alpha}^\dagger [H_{\mathrm{full}}(\mathbf{U})]_{\mathbf{n},\mathbf{n}'}^{\alpha\beta} \hat{\psi}_{\mathbf{n}', \beta}\bigg),
\end{equation}
and its matrix representation is
\begin{equation}
	U_{\mathrm{corr}}(\mathbf{U})=e^{iH_{\mathrm{full}}(\mathbf{U})}.
\end{equation}
The fermionic projector used in the Monte Carlo calculation is therefore
\begin{equation}
	P(\mathbf{U})=U_{\mathrm{corr}}(\mathbf{U})\,P_N\,U_{\mathrm{corr}}^\dagger(\mathbf{U}).
\end{equation}

\subsection{Gauge covariance, Gauss's law, and parameter counting}
\label{sec:gauge_covariance}

Gauge covariance follows from the covariance of the ingredients. 
If $G(\Omega)=\bigoplus_\mathbf{n}\Omega_\mathbf{n}$ is the matrix representation of a local gauge transformation and we denote the transformed gauge configurations as $\mathbf{U}^{G(\Omega)}$, then the following equalities hold
\begin{align}
	h_{\mathrm{MH}}(\mathbf{U}^{G(\Omega)}) & = G(\Omega)\,h_{\mathrm{MH}}(\mathbf{U})\,G^\dagger(\Omega), \\
	W_d(\mathbf{U}^{G(\Omega)})             & =G(\Omega)\,W_d(\mathbf{U})\,G^\dagger(\Omega).
\end{align}
Accordingly,
\begin{align}
	V_{\mathrm{occ}} (\mathbf{U}^{G(\Omega)})  & =G(\Omega)V_{\mathrm{occ}}(\mathbf{U})R_{\mathrm{occ}},     \\
	V_{\mathrm{unocc}} (\mathbf{U}^{G(\Omega)})& = G(\Omega)V_{\mathrm{unocc}}(\mathbf{U})R_{\mathrm{unocc}},
\end{align}
where $R_{\mathrm{occ}}$ and $R_{\mathrm{unocc}}$ are block-diagonal matrices containing arbitrary $U(2)$ rotations inside the Kramers pairs. 
Since the coefficient matrices are scalars on each $2\times2$ pair block,
\begin{equation}
	A(\mathbf{U}^{G(\Omega)})= R_{\mathrm{occ}}^\dagger A(\mathbf{U})R_{\mathrm{unocc}},
\end{equation}
and therefore
\begin{equation}
	H_{\mathrm{full}}(\mathbf{U}^{G(\Omega)})= G(\Omega)\,H_{\mathrm{full}}(\mathbf{U})\,G^\dagger(\Omega).
\end{equation}
Because $P_N$ is gauge invariant, this implies
\begin{equation}
	P(\mathbf{U}^{G(\Omega)})=G(\Omega)\,P(\mathbf{U})\,G^\dagger(\Omega).
\end{equation}
Thus the fermionic occupation matrix transforms covariantly. 
Any physical observable is a trace over the color space and site indices of the fermion occupation matrix or products between it and other gauge-covariant objects. 
Using the above transformation it is easy to see that $\hat{U}_\mathrm{corr}(\mathbf{U})$ is gauge covariant. 
Combining it with the gauge-invariant $\Psi_G(\mathbf{U})$ and with the gauge-invariant Haar measure $\mathcal D\mathbf U$, yields a gauge-invariant total ansatz $\ket{\Psi}$. 

At half filling, $N_f=L^2$, so $n_{\mathrm{occ}}=n_{\mathrm{unocc}}=L^2$, the occupied and unoccupied sectors contain $L^2/2$ Kramers pairs each. 
With $9$ channels and one complex coefficient per occupied-pair/unoccupied-pair block, the ansatz contains
\begin{equation}
	9\left(\frac{L^2}{2}\right)\left(\frac{L^2}{2}\right)=\frac{9}{4}L^4
\end{equation}
complex parameters, equivalent to $\frac{9}{2}L^4$ real variational parameters. 
However, there was no attempt to enforce additional physical symmetries of the matter sector on these parameters, meaning the number of independent parameters might be smaller than this count.

\begin{figure}[t]
	\centering
	\includegraphics[width=1.0\linewidth]{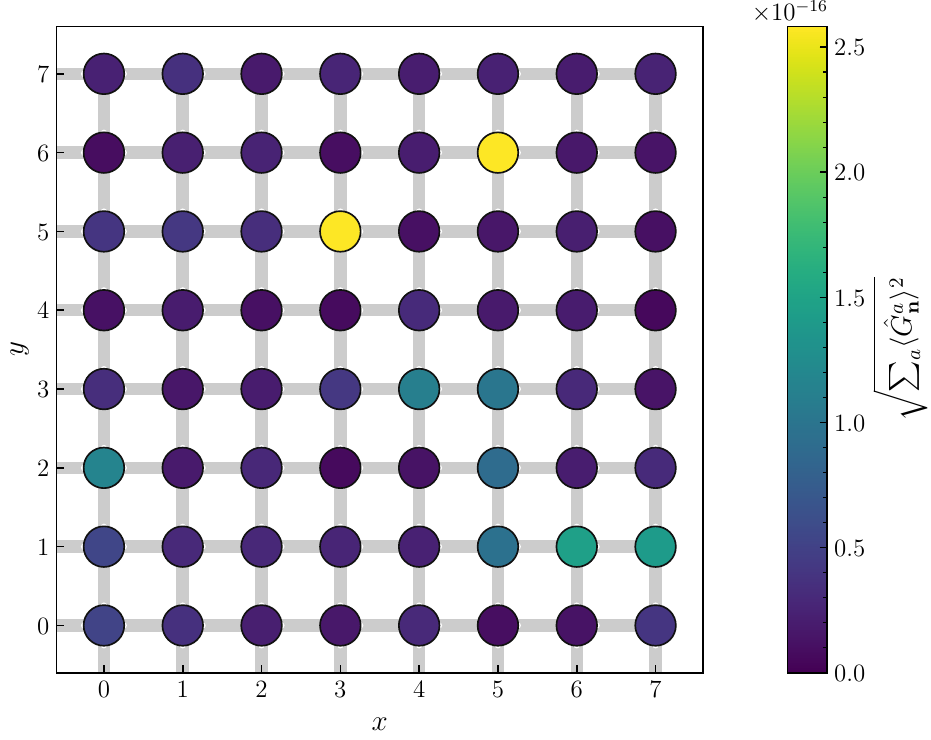}
	\caption{Gauss-law diagnostic $\sqrt{\sum_a\langle \hat{G}^a_\mathbf{n}\rangle^2}$ for an $L=8$ lattice at $(g^2,t,m)=(\sqrt{8},1.0,0.5)$ and $\lambda=\frac{4}{g^2}$. 
    The values remain at the level of machine precision, which is consistent with the gauge-covariant construction of the ansatz. 
    This plot is a numerical consistency check rather than a formal proof of gauge invariance.}
	\label{fig:GaussLawTest8}
\end{figure}

As an additional numerical consistency check, we evaluate the local Gauss-law diagnostic $\sqrt{\sum_a\langle \hat{G}^a_\mathbf{n}\rangle^2}$ for trained parameters. 
For a run at $(g^2,t,m)=(\sqrt{8},1.0,0.5)$ along the physical line $\lambda=\frac{4}{g^2}$ on an $8\times8$ lattice, we observe agreement of the Gauss's law at every vertex up to machine precision (see  Fig.~\ref{fig:GaussLawTest8}).
The gauge invariance was also verified up to machine precision for different sizes and parameter regimes. 
The plot confirms that the code implementation is consistent with the theoretically predicted gauge invariance.

\section{Analytical expectation values and training objective}
\label{sec:analytical_expressions}
When constructing a variational ansatz, we not only have to fulfill the physical constraints like gauge invariance, but also have to demonstrate the efficient numerical evaluation of observables of interest.
To derive the Monte Carlo estimators, let us first describe again the structure of the ansatz used in this work, as it will be important in deriving analytical formulas. 
The full variational state is
\begin{equation}
	\ket{\Psi}=\int \mathcal{D}\mathbf{U}\, \Psi_G(\mathbf{U})\, \hat{U}_{\mathrm{corr}}(\mathbf{U})\ket{\Psi_N}\ket{\mathbf{U}},
    \label{eqn:TotalState}
\end{equation}
where $\Psi_G(\mathbf{U})$ is a gauge-invariant wavefunction, and $\hat{U}_{\mathrm{corr}}(\mathbf{U})$ is the Gaussian unitary defined in Eq.~\eqref{eqn:UGSdefinition}. 
It is convenient to denote by $\hat{\boldsymbol{\psi}}$ the column vector collecting all fermionic annihilation operators $\hat{\psi}_{\mathbf{n}, \alpha}$. 
The simplest term to evaluate is $\hat H_B$: in the magnetic basis $\ket{\mathbf U}$ it is diagonal, so it is simply obtained as
\begin{equation}
	\langle \hat{H}_B \rangle =\int \mathcal{D}\mathbf{U}\, \vert\Psi_G(\mathbf{U})\vert^2 H_B(\mathbf{U}),
\end{equation}
where one uses Monte Carlo to sample from $p(\mathbf{U}) \equiv \vert\Psi_G(\mathbf{U})\vert^2$ and compute $\langle \hat{H}_B \rangle$. 
The same Gaussian structure also makes it simple to compute the expectation value of the quadratic matter operator $\hat{H}_{\mathrm{MH}}(\mathbf{U})=\hat{\boldsymbol{\psi}}^\dagger h_{\mathrm{MH}}(\mathbf{U}) \hat{\boldsymbol{\psi}}$, as follows
\begin{align}
		 & \langle \hat{H}_{\mathrm{MH}} \rangle = \int \mathcal{D}\mathbf{U}\mathcal{D}\mathbf{U'} \Psi_G^*(\mathbf{U'})\Psi_G(\mathbf{U}) \times \nonumber                                                                                               \\
		 &  \times \bra{\Psi_N}\hat{U}^\dagger_{\mathrm{corr}}(\mathbf{U'}) \hat{\boldsymbol{\psi}}^\dagger h_{\mathrm{MH}}(\mathbf{U}) \hat{\boldsymbol{\psi}} \hat{U}_{\mathrm{corr}}(\mathbf{U})\ket{\Psi_N}\braket{\mathbf{U'}}{\mathbf{U}}=\nonumber \\
		 & =\int \mathcal{D}\mathbf{U}\vert\Psi_G(\mathbf{U})\vert^2 \times                                                                                                                                                                         \\ &\times\bra{\Psi_N}\hat{U}^\dagger_{\mathrm{corr}}(\mathbf{U}) \hat{\boldsymbol{\psi}}^\dagger h_{\mathrm{MH}}(\mathbf{U}) \hat{\boldsymbol{\psi}} \hat{U}_{\mathrm{corr}}(\mathbf{U})\ket{\Psi_N},\nonumber
\end{align}
where we used the orthogonality of basis states $\{\ket{\mathbf{U}}\}$. 
Using the matrix representation $U_{\mathrm{corr}}(\mathbf{U})=e^{iH_{\mathrm{full}}(\mathbf{U})}$, we obtain
\begin{align}
	  \langle \hat{H}_{\mathrm{MH}} \rangle &= \int \mathcal{D}\mathbf{U}p(\mathbf{U})\nonumber\Tr \left(h_{\mathrm{MH}}(\mathbf{U}) U_{\mathrm{corr}}(\mathbf{U}) P_N U_{\mathrm{corr}}^\dagger(\mathbf{U})\right) \\&\equiv\int \mathcal{D}\mathbf{U} p(\mathbf{U})\Tr \left(h_{\mathrm{MH}}(\mathbf{U}) P( \mathbf{U})\right), 
\end{align}
where $P( \mathbf{U})\equiv U_{\mathrm{corr}}(\mathbf{U}) P_N U_{\mathrm{corr}}^\dagger(\mathbf{U})$ and $p(\mathbf{U})=\vert\Psi_G(\mathbf{U})\vert^2$, to be sampled using Monte Carlo methods. 

The only remaining term of the Hamiltonian is the electric term $\hat{H}_E \propto -\frac{1}{4}\nabla^2_{S^3}$. 
This term will split in three parts, where the term where both derivatives act on $\Psi_G(\mathbf{U})$ will be exactly the same as the electric term computed in the pure gauge theory as implemented in Ref.~\cite{spriggsAccurateGroundStates2026}. 
Consequently, any details on its computations are described therein. 
We will denote the electric field term when the derivative acts twice on the gauge wavefunction as $\langle\hat{H}_E\rangle_{GG}=E_\text{elec}$. 
The terms where the derivatives act once on the fermion state will be denoted by $\langle\hat{H}_E\rangle_{\mathrm{GF}}$ and the term where both derivatives act on the fermion state is $\langle\hat{H}_E\rangle_{\mathrm{FF}}$. 
The derivations of both expectation values are in App.~\ref{app:analytderiv}. 
The two main quantities that define the contributions of the fermionic state to the expectation values are $P(\mathbf{U})$ and $f^\xi_{\mathbf{n}, \boldsymbol{\mu}_k}$. 
Using the labeling of the direction of the link as $\boldsymbol{\mu}_k$, and its origin position as $\mathbf{n}$ and a general link variable as $\xi$, we define the Hermitian operator $f^\xi_{\mathbf{n}, \boldsymbol{\mu}_k}$ as
\begin{equation}
	f^\xi_{\mathbf{n}, \boldsymbol{\mu}_k}\equiv f^\xi_{\mathbf{n}, \boldsymbol{\mu}_k}(\mathbf{U}) =\frac{1}{i}\frac{\partial U_{\mathrm{corr}}(\mathbf{U})}{\partial \xi} U_{\mathrm{corr}}^\dagger(\mathbf{U}).
\end{equation}
However, these derivatives need to be done for each of the variables parameterizing the links and for each link. This means that an $L \times L$ lattice will need $3\times 2\times L^2$ of these $f^\xi_{\mathbf{n}, \boldsymbol{\mu}_k}$ objects to be constructed.

\section{Numerical protocol and convergence diagnostics}
\label{sec:implementation_details}
In this section, we discuss how we optimize the parameters that define the total state of Eq.~\eqref{eqn:TotalState} using variational Monte Carlo, so that the state correctly represents the ground state of the Hamiltonian in Eq.~\eqref{eqn:HamiltonianLGT}.
The optimization procedure is split into two parts, one that trains the gauge wavefunction $\Psi_G(\mathbf{U})$ and one that trains the matter parameters. 
The gauge wavefunction $\Psi_G(\mathbf{U})$ is optimized with the variational Monte Carlo framework introduced in Ref.~\cite{spriggsAccurateGroundStates2026}. 
In the implementation used here, the gauge updates are carried out with stochastic reconfiguration based on the dense Jacobian quantum geometric tensor, together with a Stochastic Gradient Descent (SGD) optimizer with a cosine-decaying learning rate. 

For the matter part of the ansatz, only the terms that depend explicitly on $\hat U_{\mathrm{corr}}(\mathbf{U})$ need to be optimized. 
The matter update therefore minimizes
\begin{equation}
	\langle \hat{H}_F \rangle=\langle \hat{H}_E \rangle_{\mathrm{FF}} +\langle \hat{H}_E \rangle_{\mathrm{GF}} + \langle \hat{H}_{\mathrm{MH}} \rangle,
\end{equation}
where $\hat{H}_{\mathrm{MH}}$ is the quadratic matter operator associated with $h_{\mathrm{MH}}(\mathbf{U})$. 
The N\'eel reference state consistently yielded lower energies than a more complicated reference state such as the Born-Oppenheimer state. 
We observed that when using the Born-Oppenheimer state as the reference state, the optimizer would not improve the energy beyond it, while when starting with the N\'eel state we managed to improve on it. 
The N\'eel state is also independent of the gauge configuration, reducing the computational complexity of the model. 
Because the optimal fermionic state depends on the gauge-wavefunction distribution $\Psi_G(\mathbf{U})$, while the optimal gauge wavefunction depends in turn on the fermionic configuration, the two sectors must be trained jointly. 
During each matter-update round, $\Psi_G(\mathbf{U})$ is held fixed, so the minimization of $\langle \hat{H}_F\rangle$ is a conditional optimization problem. 
In the coupled stage, the matter parameters are updated with an AdamW optimizer together with global-norm gradient clipping. 
We consider the model converged when the total energy plateaus and the relative energy variance is small, eventually reaching zero in the limit of an exact eigenstate.

For the study of the phase diagram of the theory, the gauge wavefunction is first pre-trained for $n_{\text{pg}}$ steps at $g^2=0$ in a separate run. 
This reduces the amount of joint training required to obtain the ground state of the theory with $g^2\neq 0$. 
The state resulting from the pre-training phase is then used to warm up the matter parameters for $n_{\text{pm}}$ steps. 
A more in-depth discussion of the used optimizers, learning rates, Monte Carlo implementations, code details and the handling of the memory intensive operations is given in App.~\ref{app:NumDetailsImplementation}.

\begin{figure}[!t]
	\centering
	\includegraphics[width=\linewidth]{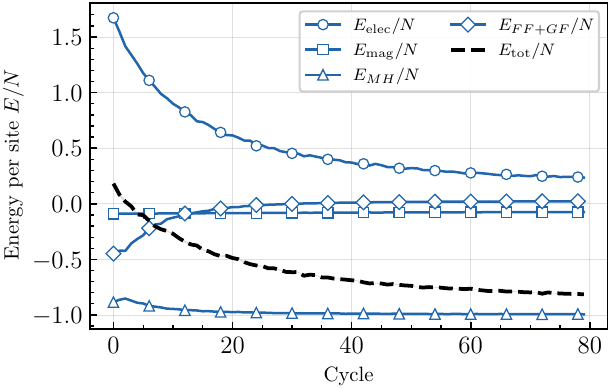}
	\vspace{0.5em}
	\includegraphics[width=\linewidth]{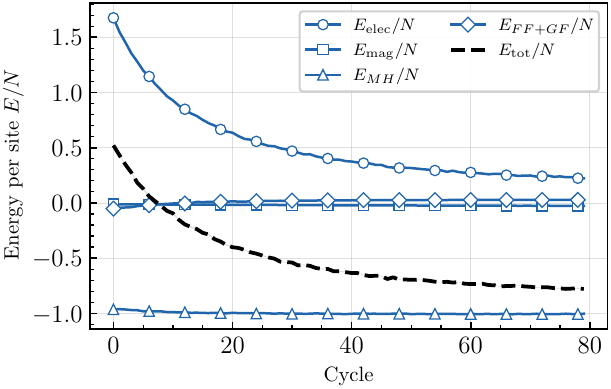}
	\caption{Cycle-by-cycle evolution of the energy contributions during joint training on an $4\times4$ lattice at $g^2=0.5$, $\lambda=-0.05$, $m=0.5$, and $t=1.0$. 
    Initialized with the pre-trained gauge wavefunction for $g^2=0$ with initial $\langle \cos \hat{B}_p\rangle=-1$ (upper panel) and with initial $\langle \cos \hat{B}_p\rangle=1$ (lower panel). 
    The decrease in $E_\text{tot}$ mostly driven by the electric energy, while $\langle \cos \hat{B}_p\rangle$ stabilizes at different values. 
    The shaded grey region around the $E_\text{tot}$ line is the Monte Carlo statistical error.}
	\label{fig:implementation_training_energy}
\end{figure}

The joint optimization is organized into $n_{\text{cycles}}$ outer cycles. 
In each cycle, we initially perform $n_{\text{ms}}$ matter-update rounds, followed by $n_{\text{gs}}$ steps of the gauge optimizer.
Within both the matter warm-up phase and the matter training rounds, for each step the matter parameters are optimized for $n_{\text{ss}}$ substeps using gauge configurations sampled from the current $\Psi_G(\mathbf{U})$. 
For each substep, we use a batch of $n_{\text{mbs}}$ samples, that are then used to compute $h_{\mathrm{MH}}(\textbf{U})$ and its occupied and unoccupied eigenvectors, $V_{\mathrm{occ}}(\textbf{U}),V_{\mathrm{unocc}}(\textbf{U})$ with a cost of $\mathcal{O}(L^6)$. 
To then compute the coefficients $f^\xi_{\mathbf{n},\boldsymbol{\mu}_k}$, we need the derivatives of the eigenvectors with respect to the link variables. 
We denote this general derivative with the simplified notation $d$. 
As used in Ref.~\cite{benderVariationalMonteCarlo2023}, one can obtain $dV_{\mathrm{occ}}(\textbf{U})$ and $dV_{\mathrm{unocc}}(\textbf{U})$ using $dh_{\mathrm{MH}}(\textbf{U})$. 
This results in $2L^2$ distinct eigenvectors with $2L^2$ components differentiated with respect to each link-variable. 
From the formulation in Eq.~\eqref{eqn:UdefinitionS3}, there are three variables per link, totalling $6L^2$ link-variables. 
This results in an $2L^2\times 6L^2\times 2L^2 = 24L^6$-size object per configuration. 
Although the object's size scales polynomially with system size, this step in the matter optimization requires a large amount of memory. 
This is compensated with lower values of $n_{\text{mbs}}$ for higher system sizes. 
We reuse the derivatives of the eigenvectors through all the $n_{\text{ss}}$ substeps, reducing the computational cost, while all the different $n_{\text{ms}}$ cycles use enough distinct samples to provide a statistically significant estimate for the optimizer to train. 
In each one of the substeps, the Wilson-basis matrices $W_d$ are also computed for each one of the $n_{\text{mbs}}$ samples. 
Their tangent tensor $dW_d$, also scaling with $L^6$, is not stored in full, since it becomes prohibitively large for bigger system sizes. 
\begin{figure}[t!]
	\centering
	\includegraphics[width=\linewidth]{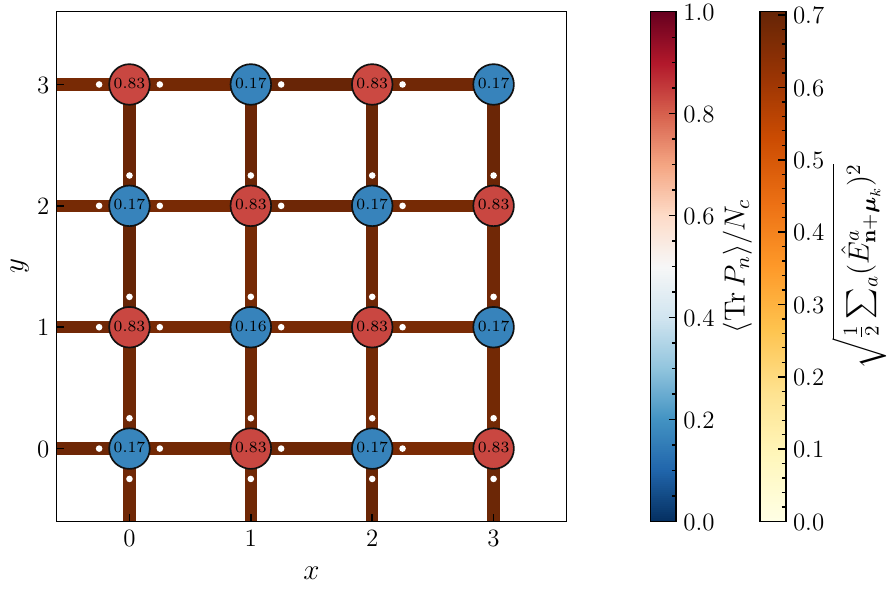}
	\caption{Spatial profile of the optimized state for the same $4\times4$ run at $g^2=0.5$, $\lambda=-0.05$, $m=0.5$, and $t=1.0$. 
    The colored circles show the site-resolved fermion occupation $\langle \Tr\,P_n\rangle/N_c$, where $P_n$ is the on-site color block of the fermionic projector $P(\mathbf{U})$ associated with lattice site $n$ and $N_c=2$ is the number of colors. 
    The link shading shows the corresponding link-resolved electric-field quantity $\sqrt{\frac{1}{2}\sum_a \langle\hat{E}_a^2\rangle}$.}
	\label{fig:implementation_occ_etotal}
\end{figure}

For the representative $4\times4$ run shown in Figs.~\ref{fig:implementation_training_energy} and~\ref{fig:implementation_occ_etotal}, we set $g^2=0.5$, $\lambda=-0.05$, $m=0.5$, and $t=1.0$, with $n_{\text{cycles}}=80$, $n_{\text{ms}}=15$, $n_{\text{ss}}=10$, $n_{\text{gs}}=200$, $n_{\text{pg}}=10000$, and $n_{\text{pm}}=20$. 
Due to the higher computational cost to train the model at higher system sizes, we chose $n_\text{cycles}=30$ for $L=6$ and $n_\text{cycles}=5$ for $L=8$. 
These values are adapted for runs on a \texttt{NVIDIA H100NVL} with \SI{94}{\giga\byte} of RAM. 
We demonstrate in Sec.~\ref{sec:physical_line} that these numbers of cycles are sufficient.
The fermionic components of the energy plotted in Fig.~\ref{fig:implementation_training_energy} only used $128$ samples for diagnostics during training as their goal is to merely indicate the general trend and not to train the model. 
Defining the individual energy components as 

\begin{align}
	E_{\mathrm{MH}} &= \langle \hat{H}_{\mathrm{MH}} \rangle, \nonumber\\
	E_{\mathrm{FF}} &= \langle \hat{H}_E \rangle_{\mathrm{FF}}, \nonumber\\
	E_{\mathrm{GF}} &= \langle \hat{H}_E \rangle_{\mathrm{GF}}, \\
	E_{\mathrm{elec}} &= \langle \hat{H}_E \rangle_{GG}, \nonumber\\
	E_{\mathrm{mag}} &= \Big\langle\lambda\sum_{\mathbf{n}} \left[1 - \frac{1}{2} \Tr \hat{P}_{\mathbf{n}, \Box}\right]\Big\rangle,\nonumber
\end{align}
and writing $N=L^2$ for the total number of sites, we obtain Fig.~\ref{fig:implementation_training_energy}, which shows the cycle-by-cycle evolution of the corresponding energy contributions during joint training using the pre-trained gauge wavefunctions for both $\langle \cos \hat{B}_p\rangle=\pm1 $. 
The initial electric contribution is comparatively large because the pre-trained gauge wavefunction is strongly peaked. 
During the joint optimization, this contribution decreases substantially and drives the total energy downward, while the magnetic term increases moderately and the fermionic contributions settle to their final values. 
The total plotted energy is obtained from summing each component of the energy from the diagnostics, while the shaded grey region around it represents the Monte Carlo error of the mean of the full training Hamiltonian, evaluated directly during the gauge optimization steps, for all the $n_\text{samples}$ samples. 
Comparing both panels, the two initializations plateau at clearly distinct energies per site, $\langle\hat H\rangle=-0.8103\pm0.0010$ and $-0.7791\pm0.0010$ (with $1024$ samples), a separation far larger than the Monte Carlo error. The gradient-based optimization is therefore trapped in two different minima.
The two states have very different values of $\langle\cos\hat{B}_p\rangle$, pointing us toward the presence of a phase transition. 
This will be discussed further in the next section.

Figure~\ref{fig:implementation_occ_etotal}, obtained using $4096$ Monte Carlo samples after training the model, shows the spatial profile of the optimized state for the same run. 
Here $P_n$ denotes the $2\times2$ on-site color block of the fermionic projector $P(\mathbf{U})$ associated with lattice site $n$, where $N_c=2$ is the number of colors. 
The site-resolved occupation $\langle \Tr\,P_n\rangle/N_c$ shows nontrivial deviations from perfect N\'eel state filling induced by the gauge-dependent Gaussian correction. 
At the same time, the link-resolved electric-field magnitude, defined as $\sqrt{\tfrac12\langle\sum_a\hat E^a_{\mathbf n,\boldsymbol\mu_k}\hat E^a_{\mathbf n,\boldsymbol\mu_k}\rangle}$, remains nearly uniform across the lattice, consistent with the translationally symmetric optimized state on periodic boundaries.

\section{Results}
\subsection{Strong-coupling benchmark}
\label{sec:high_coupling_limit}

In the strong-coupling limit where $g^2 \gg m, t$, it is possible to use perturbation theory to obtain an analytical expression for the energy of the total Hamiltonian, as a way to benchmark our algorithm. The derivation of the effective Hamiltonian is done in App.~\ref{app:StrongCoupl}. 
In the strong-coupling limit, the vacuum state $\ket{0}_\mathbf{n}$ and the color-singlet state $\frac{1}{2}\varepsilon_{\alpha\beta}\hat{\psi}^\dagger_{\mathbf{n}, \alpha}\hat{\psi}^\dagger_{\mathbf{n}, \beta}\ket{0}_\mathbf{n}\equiv \ket{2}_\mathbf{n}$ are represented by spin-states with the following correspondence: $\ket{0}\equiv\ket{\downarrow}$; $\ket{2}\equiv\ket{\uparrow}$. 
The effective Hamiltonian obtained is 
\begin{align}
	\hat{H}_\text{eff} =& \sum_\mathbf{n}m(-1)^{n_x+n_y}(1+\hat{\sigma}^z_\mathbf{n}) + \nonumber\\ &+ \frac{2t^2}{3g^2}\sum_\mathbf{n, \boldsymbol{\mu}_k}\left(\hat{\boldsymbol{\sigma}}_\mathbf{n}\cdot\hat{\boldsymbol{\sigma}}_{\mathbf{n}+\boldsymbol{\mu}_k} +\hat{\sigma}_\mathbf{n}^z\hat{\sigma}_{\mathbf{n}+\boldsymbol{\mu}_k}^z - 2\right),
	\label{eqn:TotalEffHam}
\end{align}
where $\hat{\boldsymbol{\sigma}}_\mathbf{n}\cdot\hat{\boldsymbol{\sigma}}_{\mathbf{n}+\boldsymbol{\mu}_k}=\hat{\sigma}_\mathbf{n}^x\hat{\sigma}_{\mathbf{n}+\boldsymbol{\mu}_k}^x+\hat{\sigma}_\mathbf{n}^y\hat{\sigma}_{\mathbf{n}+\boldsymbol{\mu}_k}^y+\hat{\sigma}_\mathbf{n}^z\hat{\sigma}_{\mathbf{n}+\boldsymbol{\mu}_k}^z$. 
To benchmark the variational algorithm in this regime, we compare the perturbative prediction for the ground state energy with the optimized energy returned by the algorithm for selected parameter values deep in the strong-coupling region. 
Table~\ref{tab:high_coupling_benchmark_tab} demonstrates the excellent agreement between the variational method and perturbation theory on several benchmark points.
For the case $t=m=0.5$, where the agreement is worse, it is not necessarily a sign for a worse performance of the variational algorithm, but might be because the first-order correction of $\mathcal{O}(m)$ is smaller, meaning that the neglected higher-order corrections can have a larger relative effect. 
In summary, all differences are inside the Monte Carlo error bars.

\begin{table}[ht]
	\centering
	\caption{Benchmark for comparing perturbation-theory and variational ground state energies in the strong-coupling regime for $g^2=20.0$, $\lambda=\frac{4}{g^2}$ and $L=4$ with the constant $N\lambda$ of $\hat H_B$ omitted from both energies. 
    Here $e_{\mathrm{PT}}$ denotes the perturbative prediction from the strong-coupling expansion of Eq.~\eqref{eqn:TotalEffHam}, obtained from exact diagonalization, while $e_{\mathrm{alg}}=E/N$ is the energy obtained from the variational algorithm, normalized over the number of lattice sites.}
	\begin{tabular}{ccccc}
		\hline\hline
		$t$ & $m$ & $e_{\mathrm{PT}}$ & $e_{\mathrm{alg}}$ & $\frac{|e_{\mathrm{alg}}-e_{\mathrm{PT}}|}{|e_{\mathrm{PT}}|}$ \\
		\hline
		0.5 & 1.0 & -1.0668 & -1.0626 $\pm$ 0.0053 & 0.4\% \\
		0.2 & 1.0 & -1.0107 & -1.0104 $\pm$ 0.0009 & 0.03\% \\
		0.1 & 1.0 & -1.0027 & -1.0026 $\pm$ 0.0002& 0.02\% \\
		0.5 & 0.5 & -0.5669 & -0.5756 $\pm$ 0.0095 & 1.5\% \\

		\hline\hline
	\end{tabular}
    \label{tab:high_coupling_benchmark_tab}
\end{table}

\subsection{Extended ground state phase diagram in $(g^2,\lambda)$}
\label{sec:phase_diagram}

\begin{figure}[!t]
	\centering
	\includegraphics[width=1.0\linewidth]{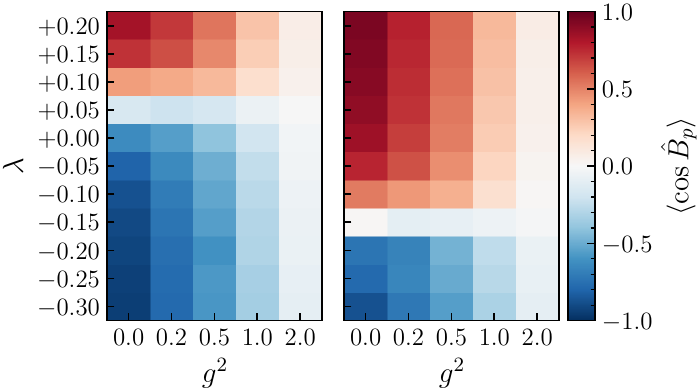}
	\caption{Color maps of the plaquette observable $\langle \cos \hat B_p\rangle$ on the $(g^2,\lambda)$ grid for $L=4$. 
    The horizontal axis is $g^2$, the vertical axis is $\lambda$, and the color scale gives $\langle \cos \hat B_p\rangle$. 
    The left panel corresponds to the initialization with $\langle \cos \hat B_p\rangle=-1$, while the right panel corresponds to the $\langle \cos \hat B_p\rangle=+1$ initialization. 
    The Monte Carlo error is less than $1\%$ at every point.}
    \label{fig:cosBp_grid}
\end{figure}

When $\lambda$ and $g^2$ are kept independent of each other in the Hamiltonian of Eq.~\eqref{eqn:HamiltonianLGT}, one obtains a behavior consistent with a phase transition between a magnetic phase ordered with $\langle\cos\hat{B}_p\rangle=1$ for $\lambda >0$ and a phase with $\langle\cos\hat{B}_p\rangle=-1$ for $\lambda \ll 0$, with a transition at $\lambda^\ast \sim -0.04$. 
The full in depth study of this phase transition and its physical implications is presented in the companion paper~\cite{rouxinolMachineLearningBased2026}. 
As shown above in the training curves, the choice of the initial state can be decisive in determining what the final state of the algorithm is, even resulting in a wrong determination of the ground state of the system. 
The analysis done above can be extended for other values of $(g^2, \lambda)$ resulting in a $2$D grid of values of $\langle \cos \hat B_p\rangle$ for different initial states. 
For the $g^2=0.0$ run, the $\langle \cos \hat B_p\rangle=1$ initialization is done by setting the $\alpha$ parameter in the Jastrow ansatz as defined in Ref.~\cite{spriggsAccurateGroundStates2026} to be a large positive value, while for $\langle \cos \hat B_p\rangle=-1$ the inverse is done. 
The model is then trained for a certain $\lambda$ and the obtained result is then used to initialize all $g^2\neq 0.0$ runs. 
The grid obtained from this procedure is plotted in Fig.~\ref{fig:cosBp_grid}. 
In the region of $\lambda \in [-0.15, 0.05]$, we observe that the resulting values of $\langle \cos \hat B_p\rangle$ are completely different between the two initializations. 
As shown by the training curves for $\lambda=-0.05$, this is not simply a result of a low number of training steps or a lack of expressivity of the ansatz, but an optimization problem due to the presence of a phase transition. 
Due to the presence of this transition between two magnetic phases, the state initialized with a very low value of $\langle \cos \hat B_p\rangle$, even for $\lambda >0$, where  $\langle \cos \hat B_p\rangle=1$ is expected to be preferred, predicts $\langle \cos \hat B_p\rangle<0$. 
However, this is easy to explain when considering that for slightly lower $\lambda$, $\langle \cos \hat B_p\rangle<0$ was actually the preferred final state of the system. 
This means that there is a local minimum of the energy which needs to be surpassed by the optimizer. 
However, to do that, it needs to cross high-energy states to reach the ground state, something which the optimizer does not favor. 
This problem needs to be considered when choosing the initialization, but it can also be used to detect phase transitions~\cite{creutzMonteCarloStudy1979}.

\begin{figure}[!t]
	\centering
	\includegraphics[width=1.0\linewidth]{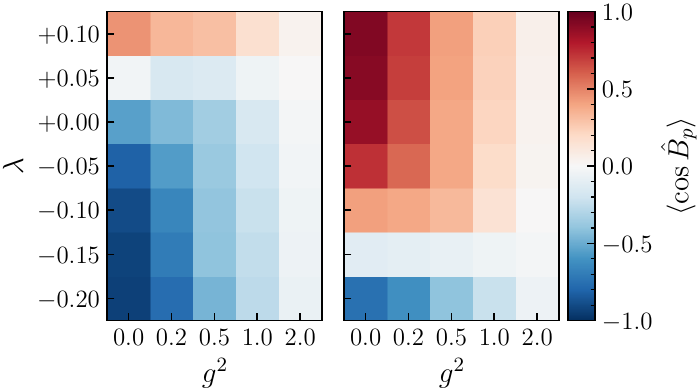}
	\caption{Color maps of the plaquette observable $\langle \cos \hat B_p\rangle$ on the $(g^2,\lambda)$ grid for $L=6$. 
    The region studied is smaller than in Fig.~\ref{fig:cosBp_grid}, due to computational constraints. 
    The labeling and the obtained conclusions are similar to $L=4$. 
    The left panel corresponds to the initialization with $\langle \cos \hat B_p\rangle=-1$, while the right panel corresponds to the $\langle \cos \hat B_p\rangle=+1$ initialization. The Monte Carlo error is less than $1\%$ at every point.}
	\label{fig:cosBp_grid_L6}
\end{figure}

To avoid these types of problems, the gauge wavefunction is initialized as the ground state of the pure gauge Hamiltonian, and the matter pre-training done on that initialization, in order to start training already close to the ground state of the total system. 
To then train larger system sizes using the results from smaller system sizes, we use the translationally invariant construction of the gauge wavefunction as described in Ref.~\cite{spriggsAccurateGroundStates2026}, which allows us to start the more expensive larger systems close to their ground state. 
This is seen in the plot of the magnetic transition for $L=6$ of Fig.~\ref{fig:cosBp_grid_L6}, where the magnetic transition scan shown in Fig.~\ref{fig:cosBp_grid} is used as the initialization of the wavefunction at $L=6$. 
To initialize the $L=6$ gauge wavefunction, the gauge parameters obtained after training at $L=4$ for each $(g^2, \lambda)$ are loaded into the $L=6$ model for the same $(g^2, \lambda)$. 
At this larger system size, the overall qualitative behavior remains the same and the transition point is consistent with the $L=4$ value within our resolution.

\begin{figure}[ht]
\begin{minipage}[t]{\linewidth}
    \centering
    \includegraphics[width=\linewidth]{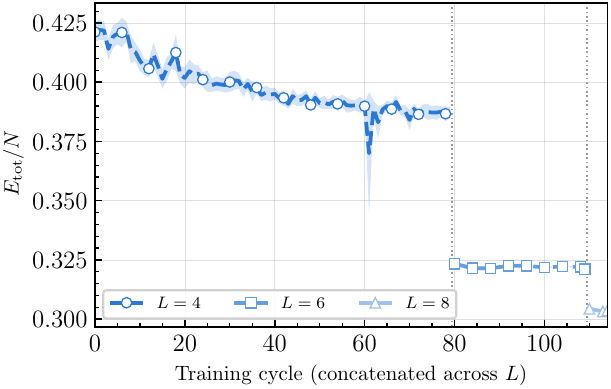}
    \vspace{-10mm}
\end{minipage}\hfill
\begin{minipage}[t]{\linewidth}
    \centering
    \includegraphics[width=\linewidth]{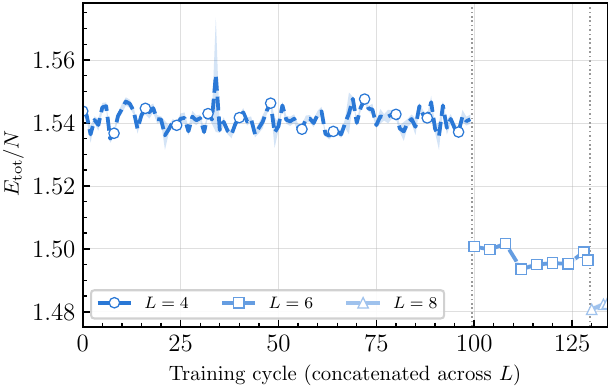}
    \vspace{-10mm}
\end{minipage}
\caption{Training of the total energy of the system as the size $L$ increases for $g^2=\sqrt{8}\approx 2.828$ (upper panel) and $g^2=0.2\sqrt{10}\approx 0.632$ (lower panel). 
The model is initialized with the pre-trained pure gauge wavefunction for $L=4$ as the starting point for the total Hamiltonian training. 
After the ground state for $L=4$ is obtained, we use the trained gauge parameters to initialize the gauge wavefunction at $L=6$ and then successively from $L=6$ to $L=8$. 
Before the joint training between the gauge and matter sector, there is a matter warm-up using the loaded gauge parameters. 
That is responsible for the reduction of the total energy seen around the vertical dashed lines, when we change system sizes. 
For the case of $g^2\approx 0.632$ the training for $L=4$ was instead done with $n_\text{cycles}=100$.}
\label{fig:trainingLincreases}
\end{figure}

\subsection{Study along the physical line}
\label{sec:physical_line}

Following the previous study, we then set $\lambda=\frac{4}{g^2}$, the correct LGT relation, and study how the total energy of the system evolves along the training cycles and for different system sizes $L$, as plotted in Fig.~\ref{fig:trainingLincreases}. 
Firstly, we note that for smaller $g^2$, the gauge pre-training together with the matter warm-up is sufficient to bring the system close to its ground state, consistent with the near-classical gauge background at small $g^2$. 
Secondly, at each size transition the energy drops sharply as the larger system is warm-started from the smaller one, and the subsequent joint-training cycles produce only small further changes. 
We read this as evidence that transferring the smaller-$L$ parameters yields a good initialization for the larger system. 
As the matter sector is re-optimized for only $n_{\text{cycles}}=30$ ($L=6$) and $5$ ($L=8$) cycles, the near-flat plateaus reflect the quality of the transfer of the gauge parameters and of the matter parameters warm-up. 
This means we achieved a good representation of the ground state of the total system. No larger lattices were studied, owing to memory constraints. 

\begin{figure}[!t]
    \centering
    \includegraphics[width=\linewidth]{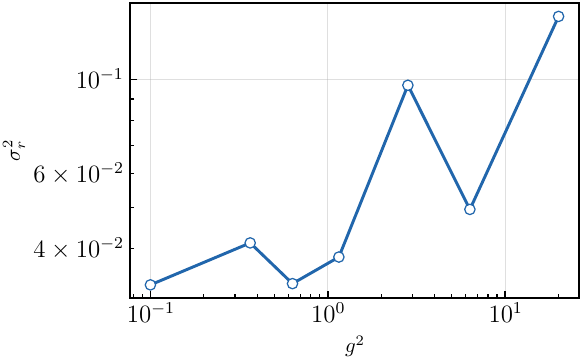}
    \caption{Evolution of the relative variance $\sigma^2_r$ with the electric coupling $g^2$ for the obtained variational state for an $8\times 8$ lattice with $t=1.0$ and $m=0.5$. 
    For all the couplings, its value follows $\sigma^2_r\sim 0.01$-$0.1$, showing that we have a good approximation of the ground state of the theory.}
    \label{fig:RelVariance}
\end{figure}

Another way to assess how well our ansatz approximates the ground state is to check the value of the relative variance~\cite{wuVariationalBenchmarksQuantum2024}, defined as $\sigma^2_r=\frac{\sigma^2}{E^2}N_\text{edges}$, where $N_\text{edges}$ is the number of edges in the lattice and $\sigma^2$ is the variance of the energy obtained during training, defined as 
\begin{equation}
    \sigma^2=\frac{\bra{\Psi}\hat{H}^2\ket{\Psi}}{\braket{\Psi}{\Psi}}-\left(\frac{\bra{\Psi}\hat{H}\ket{\Psi}}{\braket{\Psi}{\Psi}}\right)^2,
\end{equation}
which is zero in the case of $\ket{\Psi}$ being an eigenstate. 
The relative variance is then a dimensionless, intensive quantity: the smaller it is, the closer the state is to an eigenstate. 
Plotting it for the used values of $g^2$ for the $8\times 8$ lattice, we obtain the results in Fig.~\ref{fig:RelVariance}. 
There we see that the relative variance has values around $\sigma_r^2\sim 0.01$-$0.1$, showing we obtain a good approximation, which has better results for low $g^2$. 
This is a natural consequence of our ansatz, as it directly captures the eigenstructure of the mass--hopping Hamiltonian, making it accurate at low $g^2$, while being less sensitive to small, local changes important at high $g^2$.
\begin{figure}[!t]
    \centering
    \includegraphics[width=1.0\linewidth]{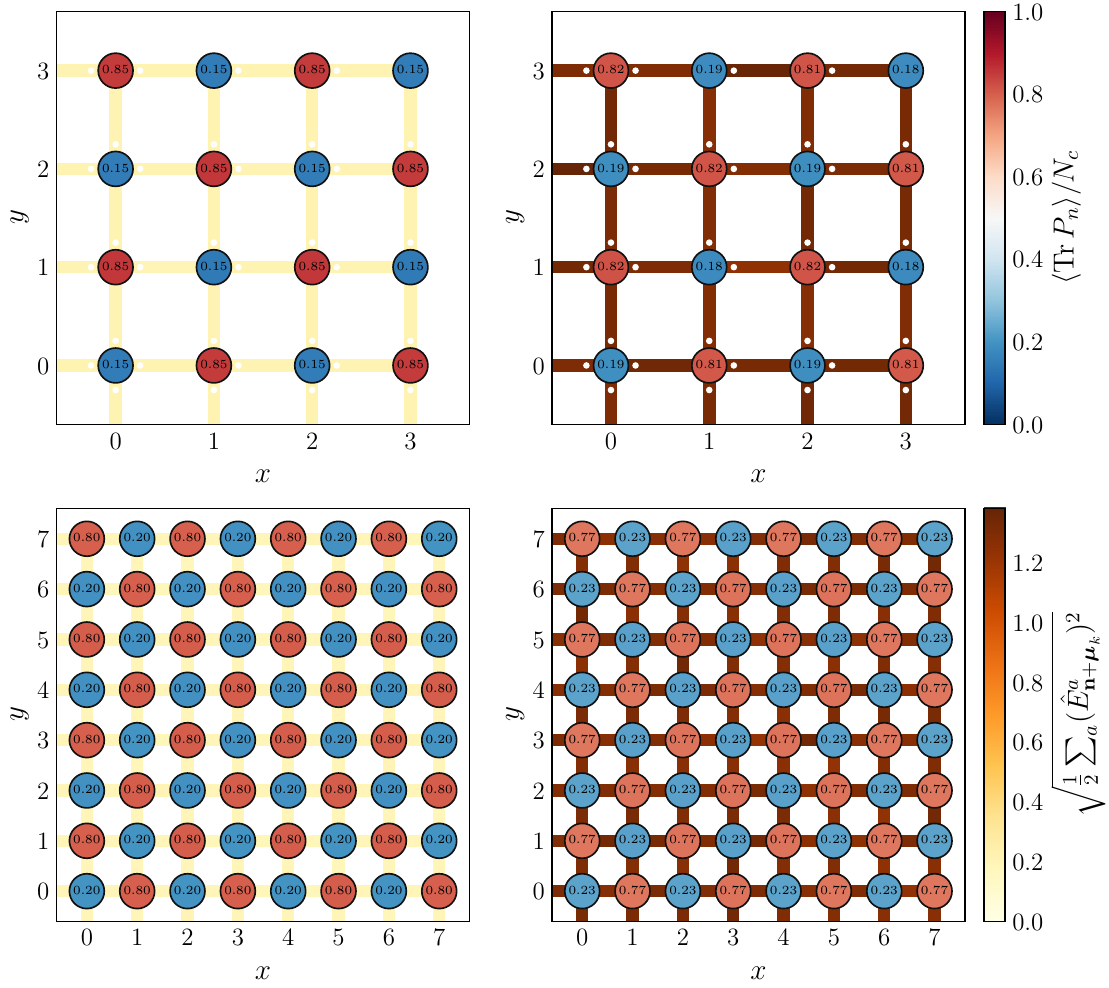}
    \caption{Spatial profile of the optimized state for the $4\times4$ (upper row) and $8\times8$ (bottom row), with $g^2\approx2.828$ (left column) and $g^2\approx0.632$ (right column), $m=0.5$, and $t=1.0$. 
    The colored circles show the site-resolved fermion occupation $\langle \Tr\,P_n\rangle/N_c$, where $P_n$ is the on-site color block of the fermionic projector $P(\mathbf{U})$ associated with lattice site $n$ and $N_c=2$ is the number of colors. 
    The link shading shows the corresponding link-resolved electric-field quantity $\sqrt{\frac{1}{2}\sum_a \langle\hat{E}_a^2\rangle}$. 
    As $g^2$ decreases the electric field increases and the state departs further away from the N\'eel state. 
    As $L$ increases, the state also departs further from a N\'eel state.}
    \label{fig:OccupPhysGrid}
\end{figure}

Apart from global quantities like the energy, our Ansatz gives us access the spatial configuration of the fermion occupation and of the electric field along the lattice. 
Using the same labeling of physical quantities as in Fig.~\ref{fig:implementation_occ_etotal}, we obtain the profiles shown in Fig.~\ref{fig:OccupPhysGrid}, with $g^2=\sqrt{8}\approx2.828$ and $g^2=0.2\sqrt{10}\approx0.632$ for $L=4$ and $L=8$ in Fig.~\ref{fig:OccupPhysGrid}. 
There we can see that decreasing $g^2$ favors hopping, moving the state away from the N\'eel state preferred for $g^2\to \infty$. 
In addition, increasing the size of the system also increases the possibility of longer-range hopping processes, which also moves the state away from the N\'eel state as $L$ is increased. 

\begin{figure}[t!]
	\centering
	\includegraphics[width=1.0\linewidth]{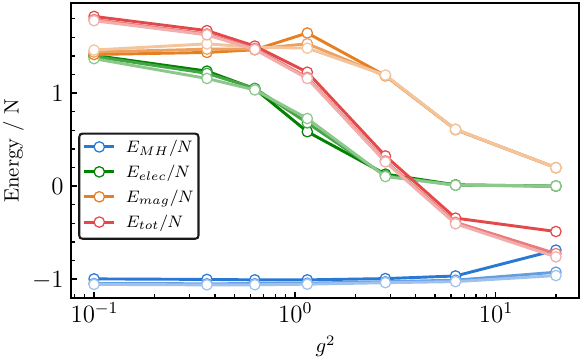}
	\caption{Evolution of the energy components, with $E_{\mathrm{MH}}=\langle \hat{H}_{\mathrm{MH}} \rangle$, $E_{\mathrm{elec}}=\langle \hat{H}_{E}\rangle_{GG}$ and $ E_{\mathrm{mag}}=\langle \hat{H}_{B}\rangle$, as a function of $g^2$ along the physical line $\lambda=\frac{4}{g^2}$, for $L=4,6$ and $8$. The lighter colors identify larger system sizes. The total energy improves as $L$ is increased, while its variation with $g^2$ is largely driven by changes in the gauge contribution.}
	\label{fig:lambdascanen}
\end{figure}

These changes can also be understood from the evolution of the different components of the Hamiltonian and of the average magnitude of the parameters of the matter ansatz. 
To understand how the total energy changes with system size, we plot the total energy and its components for different $L$ and different values of $g^2$, as shown in Fig.~\ref{fig:lambdascanen}. 
There, we see that the lighter shades, which identify larger system sizes, have a lower energy, mostly because of a decrease in the fermionic energy contributions. 
This points to an important influence of finite-size effects on the fermionic state, specially at high $g^2$. 

\section{Discussion and outlook}
\label{sec:discussion}
In this work we studied the ground state of the SU$(2)$ lattice gauge theory with dynamical fermions, by representing the fermionic state as a superposition of gauge-covariant Gaussian states, together with a machine learning-based gauge wavefunction ansatz~\cite{spriggsAccurateGroundStates2026}.
The ansatz is built from Wilson lines and the eigenvectors of the mass--hopping Hamiltonian. 
We showed that the approach satisfies Gauss's law and scales polynomially with system size.

We derived analytical tools to reduce the computational cost, mainly through simplified formulas for fermionic expectation values that exploit the Gaussian structure of the states, and validated the ansatz against strong-coupling Schrieffer-Wolff perturbation theory. 
We then described the numerical implementation in some detail, covering convergence considerations and the main practical limitations, such as memory allocation and the variational optimization getting trapped in local minima. 
For $m=0.5$ and $t=1.0$, we mapped out the ground state properties both when the magnetic coupling $\lambda$ is varied independently of the electric coupling $g^2$ and along the physical line $\lambda=\frac{4}{g^2}$, computing site- and link-resolved occupation and electric-field profiles. 
At high $g^2$, we saw that increasing $L$ lowers the mass--hopping energy and shifts the state away from the reference Néel state. 
Our results are consistent with the variational training converging to the ground state, as seen by the low relative variance and plateauing energy, and further supported by the correct $g^2\to \infty$ limit, the machine-precision Gauss-law satisfaction, and expected $L$ and $g^2$ trends.

Several natural extensions remain open. 
The Wilson lines in the matter ansatz were restricted to length two. 
Whether longer lines improve accuracy warrants investigation. 
Parametrizing the matter ansatz's variational parameters with gauge-invariant neural networks, or using a more expressive gauge wavefunction, are other routes to improve the representation of the ground state. 
On the computational side, systems larger than $L=8$ will require a refactored algorithm, as several intermediate objects become prohibitively large at that scale. 
A more efficient implementation would also allow a denser scan of $(\lambda, g^2)$ space, enabling more precise measurements of observables not studied here, such as the Wilson-loop area law or the Fredenhagen-Marcu order parameter, and therefore a sharper characterization of confinement properties. 
Additionally, enforcing global lattice symmetries (translational, rotational) in the matter ansatz would reduce the parameter count and is likely a prerequisite for extending the approach to $3+1$D.

Beyond these specific extensions, the framework serves two broader purposes: it provides a reusable foundation for variational studies of lattice gauge theories, and it offers a concrete reference for benchmarking emerging methods such as quantum simulation and tensor networks. 
This construction is a starting point for other full continuous representations of more complex gauge groups such as SU$(3)$, though the pseudoreality argument used here will require a redesign in that setting. 
It also opens a natural path toward thermal states and real-time evolution in theories with the full gauge group. 
This is, to our knowledge, the first sign-problem-free variational treatment covering the full, continuous non-Abelian, matter-coupled regime, and the methods developed here should carry over to richer gauge groups, higher dimensions, and the wider range of open questions present in these theories.

\footnotesize
\begin{acknowledgments}
	\textbf{\textit{Acknowledgments.---}} We are grateful to Jannes Nys and Thomas Spriggs for stimulating discussions.
	G.R. and J.C.H.~acknowledge funding by the Max Planck Society, the Deutsche Forschungsgemeinschaft (DFG, German Research Foundation) under Germany's Excellence Strategy - EXC-2111 - 390814868, and the European Research Council (ERC) under the European Union's Horizon Europe research and innovation program (Grant Agreement No.~101165667)-ERC Starting Grant QuSiGauge.
	This work is part of the Quantum Computing for High-Energy Physics (QC4HEP) working group.
    J.B. is supported by a Feodor Lynen Research Fellowship from the Alexander von Humboldt Foundation. 
	P.E. acknowledges the support received from the Dutch National Growth Fund (NGF) as part of the Quantum Delta NL program in the NWO-Quantum Technology program (Grant No.~NGF.1623.23.006).
	P.E. also acknowledges funding from the Carl-Zeiss-Stiftung (CZS Center QPhoton). M.G. is supported by CERN through the CERN Quantum Technology Initiative. MG thanks ESA SpaceHPC for the time provided on their infrastructure.
\end{acknowledgments}
\normalsize
\section*{Data Availability}
The data and the code used to generate all plots in this work are available in~\cite{rouxinolDataNeuralQuantum2026}.
\appendix
\section{Derivation of the analytical expression for the Hamiltonian expectation values}
\label{app:analytderiv}
In this appendix we derive the contributions of the electric term of the Hamiltonian $\hat{H}_E \propto -\frac{1}{4}\nabla^2_{S^3}$, when the derivative terms act on either once or twice in the fermionic state. This means that now the derivative can either act once on $\Psi_G(\mathbf{U})$ and once on $\hat{U}_{\mathrm{corr}}(\mathbf{U})$ or on $\hat{U}_{\mathrm{corr}}(\mathbf{U})$ twice. Firstly, we recall that
\begin{align}
	\nabla^2_{S^3} =& \frac{1}{\sin^2\!\left(\frac{\rho}{2}\right)} \Bigg( \frac{\partial}{\partial \rho}\left(4\sin^2\!\left(\frac{\rho}{2}\right)\frac{\partial}{\partial \rho}\right) + \nonumber \\
	&\frac{1}{\sin\theta}\frac{\partial}{\partial \theta}\left(\sin\theta\frac{\partial}{\partial \theta}\right) + \frac{1}{\sin^2\!\theta}\frac{\partial^2}{\partial \phi^2} \Bigg),\label{eqn:LaplaceBeltramiDefinition}
\end{align}
where the link variables $(\rho, \theta, \phi)$ change from link to link, but we drop their dependence for notational simplicity. We keep the notation where $\hat{\boldsymbol{\psi}}$  is the column vector collecting all fermionic annihilation operators $\hat{\psi}_{\mathbf{n}, \alpha}$. To determine the new contributions of $\langle\hat{H}_E\rangle$, we start by labeling an arbitrary link variable as $\xi$. 
We will need to compute
\begin{align}
	 & \frac{\partial \hat{U}_{\mathrm{corr}}(\mathbf{U})}{\partial \xi} \hat{U}^\dagger_{\mathrm{corr}}= \\
	 & =\int_0^1 ds\, i e^{is\hat{\boldsymbol{\psi}}^\dagger H_{\mathrm{full}}(\mathbf{U}) \hat{\boldsymbol{\psi}}} \hat{\boldsymbol{\psi}}^\dagger \frac{\partial H_{\mathrm{full}}(\mathbf{U})}{\partial \xi}\hat{\boldsymbol{\psi}}e^{-is\hat{\boldsymbol{\psi}}^\dagger H_{\mathrm{full}}(\mathbf{U}) \hat{\boldsymbol{\psi}}} =\nonumber \\
	 & =\hat{\boldsymbol{\psi}}^\dagger \frac{\partial U_{\mathrm{corr}}(\mathbf{U})}{\partial \xi}U_{\mathrm{corr}}^\dagger(\mathbf{U})\hat{\boldsymbol{\psi}},\nonumber
\end{align}
which satisfies $\mathcal{H}\left(\frac{\partial \hat{U}_{\mathrm{corr}}}{\partial \xi}(\mathbf{U})  \hat{U}^\dagger_{\mathrm{corr}}(\mathbf{U})\right)=\frac{\partial}{\partial \xi} \left(\hat{U}_{\mathrm{corr}}(\mathbf{U})\hat{U}^\dagger_{\mathrm{corr}}(\mathbf{U}) \right)=0$, where $\mathcal{H}(.)$ is the Hermitian part of a matrix. As a result, the components of~Eq.~\eqref{eqn:LaplaceBeltramiDefinition} with a single derivative will have no contribution, as the expectation value of an anti-Hermitian operator is purely imaginary and cannot contribute to a physical observable. 
If we consider the case where the entire Laplace-Beltrami operator acts on $\hat{U}_{\mathrm{corr}}(\mathbf{U})$, we now need to compute
\begin{align}
	 & \frac{\partial ^2\hat{U}_{\mathrm{corr}}(\mathbf{U})}{\partial \xi^ 2}=\frac{\partial }{\partial \xi} \bigg(\hat{\boldsymbol{\psi}}^\dagger \frac{\partial U_{\mathrm{corr}}(\mathbf{U})}{\partial \xi}U_{\mathrm{corr}}^\dagger(\mathbf{U})\hat{\boldsymbol{\psi}}\hat{U}_{\mathrm{corr}}\bigg) =\nonumber                                             \\
	 & \hat{\boldsymbol{\psi}}^\dagger\bigg( \frac{\partial ^2 U_{\mathrm{corr}}(\mathbf{U})}{\partial \xi^2}U_{\mathrm{corr}}^\dagger(\mathbf{U}) + \frac{\partial U_{\mathrm{corr}}(\mathbf{U})}{\partial \xi}\frac{\partial U_{\mathrm{corr}}^\dagger(\mathbf{U})}{\partial \xi}\bigg) \nonumber \\
	 & \hat{\boldsymbol{\psi}}\hat{U}_{\mathrm{corr}}(\mathbf{U}) +\hat{\boldsymbol{\psi}}^\dagger \frac{\partial U_{\mathrm{corr}}(\mathbf{U})}{\partial \xi}U_{\mathrm{corr}}^\dagger(\mathbf{U})\hat{\boldsymbol{\psi}} \times\nonumber \\
    & \times \hat{\boldsymbol{\psi}}^\dagger \frac{\partial U_{\mathrm{corr}}(\mathbf{U})}{\partial \xi}U_{\mathrm{corr}}^\dagger(\mathbf{U})\hat{\boldsymbol{\psi}}\hat{U}_{\mathrm{corr}}(\mathbf{U})  ,
\end{align}
where again the first line of the last equality is the derivative of an anti-Hermitian operator, hence anti-Hermitian itself, meaning we only need to consider the last term. 
Now recovering the labeling of the direction of the link as $\boldsymbol{\mu}_k$, and its origin position as $\mathbf{n}$, we define
\begin{equation}
	f^\xi_{\mathbf{n}, \boldsymbol{\mu}_k}\equiv f^\xi_{\mathbf{n}, \boldsymbol{\mu}_k}(\mathbf{U}) =\frac{1}{i}\frac{\partial U_{\mathrm{corr}}(\mathbf{U})}{\partial \xi} U_{\mathrm{corr}}^\dagger(\mathbf{U}),
\end{equation}
which is a Hermitian operator, as $\frac{\partial \hat{U}_{\mathrm{corr}}}{\partial \xi}(\mathbf{U})$ is anti-Hermitian. 
It represents the contribution from the electric field term $\hat{H}_E$ when it acts twice on $\hat{U}_{\mathrm{corr}}(\mathbf{U})$, giving the contribution
\begin{align}
	 & \langle \hat{H}_E\rangle_{\mathrm{FF}} =\frac{g^2}{2}\sum_{\mathbf{n}, \boldsymbol{\mu}_k}\int \mathcal{D}\mathbf{U} p(\mathbf{U})\bra{\Psi_N}\hat{U}^\dagger_{\mathrm{corr}}(\mathbf{U}) \\ &\Big( \hat{\boldsymbol{\psi}}^\dagger f^\rho_{\mathbf{n}, \boldsymbol{\mu}_k}\hat{\boldsymbol{\psi}}\hat{\boldsymbol{\psi}}^\dagger f^\rho_{\mathbf{n}, \boldsymbol{\mu}_k}\hat{\boldsymbol{\psi}}  \nonumber +\frac{1}{(2\sin\left(\frac{\rho}{2}\right))^2}\hat{\boldsymbol{\psi}}^\dagger f^\theta_{\mathbf{n}, \boldsymbol{\mu}_k}\hat{\boldsymbol{\psi}}\hat{\boldsymbol{\psi}}^\dagger f^\theta_{\mathbf{n}, \boldsymbol{\mu}_k}\hat{\boldsymbol{\psi}}+ \\
	 & +\frac{1}{(2\sin\left(\frac{\rho}{2}\right)\sin(\theta))^2}\hat{\boldsymbol{\psi}}^\dagger f^\phi_{\mathbf{n}, \boldsymbol{\mu}_k}\hat{\boldsymbol{\psi}}\hat{\boldsymbol{\psi}}^\dagger f^\phi_{\mathbf{n}, \boldsymbol{\mu}_k}\hat{\boldsymbol{\psi}} \Big)\hat{U}_{\mathrm{corr}}(\mathbf{U})\ket{\Psi_N}, \nonumber
\end{align}
where all link variables inside the sum are understood to be evaluated for the link at position $\mathbf{n} $ and direction $\boldsymbol{\mu}_k$. 
Using standard properties of the expectation values of Gaussian states, one obtains that
\begin{align}
	 & \bra{\Psi_N}\hat{U}^\dagger_{\mathrm{corr}}(\mathbf{U}) \hat{\boldsymbol{\psi}}^\dagger f^\rho_{\mathbf{n}, \boldsymbol{\mu}_k}\hat{\boldsymbol{\psi}}\hat{\boldsymbol{\psi}}^\dagger f^\rho_{\mathbf{n}, \boldsymbol{\mu}_k}\hat{\boldsymbol{\psi}} \hat{U}_{\mathrm{corr}}(\mathbf{U})\ket{\Psi_N} = \\
	 & \left[\Tr\left(P( \mathbf{U}) f^\rho_{\mathbf{n}, \boldsymbol{\mu}_k}\right)\right]^2 +\Tr\left(P( \mathbf{U}) f^\rho_{\mathbf{n}, \boldsymbol{\mu}_k}\left(\mathbb{I}-P( \mathbf{U})\right) f^\rho_{\mathbf{n}, \boldsymbol{\mu}_k}\right)\nonumber.
\end{align}
A similar procedure would yield that the term where the Laplacian acts once on the gauge wavefunction and once on the fermionic state has an expectation value $\langle \hat{H}_E \rangle_{\mathrm{GF}}$
\begin{align}
	 & \langle \hat{H}_E\rangle_{\mathrm{GF}} =\frac{g^2}{i}\sum_{\mathbf{n},k}\int \mathcal{D}\mathbf{U} p(\mathbf{U})\Bigg( \Tr\left(P( \mathbf{U}) f^\rho_{\mathbf{n}, \boldsymbol{\mu}_k}\right)\times  \\ & \frac{\partial \ln\Psi_G(\mathbf{U})}{\partial \rho_{\mathbf{n},\boldsymbol{\mu}_k}}+\frac{1}{(2\sin\left(\frac{\rho}{2}\right))^2}\Tr\left(P( \mathbf{U}) f^\theta_{\mathbf{n}, \boldsymbol{\mu}_k}\right)\frac{\partial \ln\Psi_G(\mathbf{U})}{\partial \theta_{\mathbf{n},\boldsymbol{\mu}_k}} \nonumber \\
	 & +\frac{1}{(2\sin\left(\frac{\rho}{2}\right)\sin(\theta))^2}\Tr\left(P( \mathbf{U}) f^\phi_{\mathbf{n}, \boldsymbol{\mu}_k}\right)\frac{\partial \ln\Psi_G(\mathbf{U})}{\partial \phi_{\mathbf{n},\boldsymbol{\mu}_k}} \Bigg), \nonumber
\end{align}
where the real part is understood, which reduces all the computations of the expectation values to at most computing a single derivative of $U_{\mathrm{corr}}(\mathbf{U})$ and the second order derivative of $\Psi_G(\mathbf{U})$. 

\section{Numerical implementation details}
\label{app:NumDetailsImplementation}

The code is implemented using \texttt{NetKet}~\cite{netket2:2019, netket3:2022}, which calls \texttt{JAX}~\cite{jax2018github} and \texttt{Flax}~\cite{flax2020github} as numerical backends. 
We used \texttt{NetKet}'s \texttt{QGTJacobianDense} function with a \texttt{diag\_shift}$=1\times 10^{-3}$, \texttt{holomorphic=False}, and \texttt{optax.sgd} with a cosine decay from $10^{-3}$ to $3\times10^{-4}$. 
Relative to the pure-gauge case, the only modification is that the cost function now includes the fermionic contributions to the energy. 
The overall structure of the gauge optimization is otherwise unchanged, so the scaling discussion of Ref.~\cite{spriggsAccurateGroundStates2026} still applies for the pure gauge Hamiltonian. In all of our work, we took the reference state for the matter sector to be the half-filled N\'eel state. This is done in the code by selecting the \texttt{high} initialization option. Another option is to test the Born-Oppenheimer ansatz as the reference state, selectable via the \texttt{low} option. In the matter warm-up phase, the optimizer used is \texttt{AdamW} with clip norm $10.0$, a learning rate cosine schedule with initial value $0.01$ and a weight decay of $10^{-5}$. In the matter training inside the main training cycles, we use an \texttt{AdamW} optimizer with clip norm $0.1$, weight decay $10^{-5}$, and learning rate $5\times10^{-4}\times 0.1^{c/(n_{\text{cycles}}-1)}$, where $c=0,1,\ldots, n_{\text{cycles}}-1$ counts the current cycle. 

All expectation values needed to train the model are evaluated by Monte Carlo using $1024$ samples, with $64$ chains and $32$ thermalization samples per chain. The sampler also computes the integrated autocorrelation time $\tau$ and the Gelman--Rubin statistic $\hat{R}$~\cite{spriggsAccurateGroundStates2026,gelmanInferenceIterativeSimulation1992} 
during training. The autocorrelation time is measured in units of sampler steps, where each step sweeps across all lattice edges, and the Gelman--Rubin statistic is measured over the $64$ chains. Across all runs we find $\tau\lessapprox 1$, a sign of independent samples, and $\hat{R}\lessapprox 1.05$, indicating
well-mixed chains with no detectable departure from a common stationary
distribution. 
The effective sample size is therefore close to the number of samples, and we quote the Monte Carlo error as the standard error of the mean, without an autocorrelation correction.

To compute all objects inside our memory constraints, we used a chunk size of $32$ for $L=4,6$, and of $2$ for $L=8$, which made the training much slower. Each one of the matter steps utilizes $1024$ samples, but for each one of the $n_{\text{ss}}$ substeps a batch of $256$ out of the $1024$ samples is used to evaluate the energy and its gradient. 
This number is reduced to $32$ for runs with $L=6$ and $L=8$. 
The batch-size reduction is applied for memory reasons at larger $L$. 
Its impact on the training of the fermionic part of the Hamiltonian is limited, as after the model is trained for all substeps, a set of $320=32\times10$ samples has been used. 
The code also faces a problem in storing the tangent object to the Wilson lines $dW_d$. 
Instead of computing them once and storing them, they are recomputed each time they are needed, which increases the runtime of the algorithm, but enables its execution at larger system sizes. 
However, if larger RAM GPUs were available one could avoid this bottleneck. 
These matter updates are followed by $n_{\text{gs}}$ gauge-optimization substeps. 
In the case of larger system sizes, one needs to reduce the size of each chunk that the \texttt{NetKet} optimizer computes to train the model, due to memory constraints, slowing down the algorithm.

\section{Derivation of the strong-coupling Hamiltonian}
\label{app:StrongCoupl}
To obtain an effective Hamiltonian in the strong-coupling $g^2\gg m,t$ limit we first consider the unperturbed Hamiltonian $\hat{H}_0=\frac{g^2}{2}\sum_{\mathbf{n},k,a} \hat{E}_{\mathbf{n}, \boldsymbol{\mu}_k,a} \hat{E}_{\mathbf{n}, \boldsymbol{\mu}_k,a}\equiv\frac{g^2}{2}\sum_{\mathbf{n},k} \hat{E}_{\mathbf{n}, \boldsymbol{\mu}_k}^2 \equiv \frac{g^2}{2}\hat{E}^2$. 
A standard way to represent the eigenstates of $\hat{E}_{\mathbf{n}, \boldsymbol{\mu}_k}^2 =\hat{E}_{\mathbf{n}, \boldsymbol{\mu}_k,L}^2=\hat{E}_{\mathbf{n}, \boldsymbol{\mu}_k, R}^2$ is by representing them as an irreducible representation of SU$(2)$ in an equivalent way to quantum angular momentum spinors $\ket{j,m_L, m_R}$, where $j$, defined such that $\hat{E}_{\mathbf{n}, \boldsymbol{\mu}_k}^2\ket{j,m_L, m_R} = j(j+1)\ket{j,m_L, m_R}$, quantifies the flux on the link and $m_{L/R}$ are the eigenvalues of $\hat{E}^{(3)}_{\mathbf{n}, \boldsymbol{\mu}_k,L/R}$. 
By defining our gauge basis using $\ket{j,m_L, m_R}$, it is clear that the ground state of $\hat{H}_0$ is $\ket{0, 0, 0}$. 
Automatically, Gauss's law requires $\hat\rho^a_\mathbf{n}\ket{\text{phys}}=0$ at every site, so each site must carry vanishing color charge. The only such on-site states are $\ket 0$, annihilated by all $\hat\psi_{\mathbf n,\alpha}$, and the color singlet $\frac{1}{2}\varepsilon_{\alpha\beta}\hat{\psi}^\dagger_{\mathbf{n}, \alpha}\hat{\psi}^\dagger_{\mathbf{n}, \beta}\ket{0}_\mathbf{n}\equiv \ket{2}_\mathbf{n}$.
Thus, the ground state of $\hat{H}_0$ is a highly degenerate space spanned by $2^N$ states, where $N$ is the number of lattice sites. 
We then remain with the perturbation Hamiltonian
\begin{align}
	 & \hat{V}=\hat{H}_B+\hat{H}_m+\hat{H}_t =                                                                       \label{eqn:PotentialPerturbation}                                                                                  \\
	 & \lambda\sum_{\mathbf{n}} \left[1 - \frac{1}{2} \Tr \hat{P}_{\mathbf{n}, \Box}\right]  +\sum_{\mathbf{n}, \alpha}  m(-1)^{n_x+n_y}\hat{\psi}_{\mathbf{n}, \alpha}^\dagger\hat{\psi}_{\mathbf{n}, \alpha}  \nonumber 
	\\&-\frac{it}{2} \sum_{\mathbf{n}, \boldsymbol{\mu}_k, \alpha, \beta} \left(\hat{\psi}_{\mathbf{n}, \alpha}^\dagger \hat{U}^{\alpha\beta}_{\mathbf{n},\boldsymbol{\mu}_k} \hat{\psi}_{\mathbf{n}+\boldsymbol{\mu}_k, \beta} \eta_{\mathbf{n},\boldsymbol{\mu}_k} - \text{H.c.} \nonumber\right),
\end{align}
which we can use together with the Schrieffer-Wolff transformation~\cite{schriefferRelationAndersonKondo1966} to build an effective Hamiltonian for the $g^2\to \infty$ limit. We will neglect the additive constant of $N\cdot \lambda$ from the first term of $\hat H_B$. We now define the projector onto the subspace spanned by the degenerate ground states of $\hat{H}_0=\frac{g^2}{2}\hat{E}^2$.  
For simplicity, let us reduce our discussion to a single link and the sites at its ends, labelled by $+, -$, the signs they obtain due to the $(-1)^{n_x+n_y}$ staggering in the mass term. The basis to represent a specific link then follows $\ket{j, m_L, m_R}\otimes\ket{a}_+\otimes\ket{b}_-$, where $a,b=0,\alpha,2$ and $\ket{\alpha}_\mathbf{n}=\hat{\psi}^\dagger_{\mathbf{n},\alpha}\ket{0}$. The part of the projector that acts on that link is
\begin{align}
	\hat{P}_l= & \ket{0,0,0}\bra{0,0,0}\otimes                                                                   \\
	     & (\ket{0}_+\bra{0}_++\ket{2}_+\bra{2}_+)\otimes(\ket{0}_-\bra{0}_-+\ket{2}_-\bra{2}_-)\nonumber,
\end{align}
which we can use to define the first order effective Hamiltonian
\begin{equation}
	\hat{H}_\text{eff}^{(1)}=\hat{P}_l\hat{V}\hat{P}_l=\hat{P}_l\hat{H}_m\hat{P}_l=2m(\ket{2}_+\bra{2}_+-\ket{2}_-\bra{2}_-),
\end{equation}
where the terms $\hat{H}_t$ and $\hat{H}_B$ create transitions from the space projected by $\hat{P}_l$ to $\mathbb{I}-\hat{P}_l$, meaning they yield a zero contribution in first order. 
If we map the matter states $\ket{0}_\mathbf{n}$ and $\ket{2}_\mathbf{n}$ to spin-states of Pauli operators, $\ket{0}\equiv\ket{\downarrow}$ and $\ket{2}\equiv\ket{\uparrow}$, we obtain
\begin{equation}
	\hat{H}_\text{eff}^{(1)} = \sum_\mathbf{n}m(-1)^{n_x+n_y}(1+\hat{\sigma}^z_\mathbf{n}).
\end{equation}
If we now go to second order in perturbation theory, we need to consider the contribution
\begin{equation}
	\hat{H}_\text{eff}^{(2)}=-\hat{P}_l\hat{V}\frac{1-\hat{P}_l}{\hat{H}_0-E_0}\hat{V}\hat{P}_l,
\end{equation}
whose main contribution arises when $\hat{V}\approx \hat{H}_t$, as any extra contribution from $\hat{H}_B$ would yield an extra multiplicative coefficient $\lambda/2$. In the physical theory this scales as $1/g^2$, and is therefore subleading. 
To do that, first consider the formula of $\hat{H}_t$ at each link with direction $\mu$
\begin{align}
	\hat{H}_{t,l} & =-\frac{it}{2} \sum_{\mathbf{n}, \boldsymbol{\mu}_k} \left(\hat{\psi}_{\mathbf{n}, \alpha}^\dagger \hat{U}^{\alpha\beta}_{\mathbf{n},\boldsymbol{\mu}_k} \hat{\psi}_{\mathbf{n}+\boldsymbol{\mu}_k, \beta} \eta_{\mathbf{n},\boldsymbol{\mu}_k} - \text{H.c.} \nonumber\right)\\&=-\frac{it}{2}\left(\hat{A}-\hat{A}^\dagger\right),
\end{align}
where we can see that $\hat{A}\ket{0}_{\mathbf{n}+\boldsymbol{\mu}_k}=\hat{A}\ket{2}_\mathbf{n}=0$ and $\hat{A}^\dagger\ket{2}_{\mathbf{n}+\boldsymbol{\mu}_k} =\hat{A}^\dagger\ket{0}_\mathbf{n} =0$. 
This means that
\begin{align}
	 & \hat{H}_{t,l}\hat{P}_l=-\frac{it}{2}\bigg(\hat{A}\ket{0,0,0}\ket{0}_\mathbf{n}\ket{2}_{\mathbf{n}+\boldsymbol{\mu}_k}\bra{0,0,0}\bra{0}_\mathbf{n}\bra{2}_{\mathbf{n}+\boldsymbol{\mu}_k}-\nonumber \\&\hat{A}^\dagger\ket{0,0,0}\ket{0}_{\mathbf{n}+\boldsymbol{\mu}_k}\ket{2}_\mathbf{n}\bra{0,0,0}\bra{0}_{\mathbf{n}+\boldsymbol{\mu}_k}\bra{2}_\mathbf{n}\bigg),
\end{align}
where we now compute
\onecolumngrid
\begin{align}
        \hat{A}\ket{0,0,0}\ket{0}_\mathbf{n}\ket{2}_{\mathbf{n}+\boldsymbol{\mu}_k} & =\hat{\psi}_{\mathbf{n},\alpha}^\dagger \hat{U}^{\alpha\beta}_{\mathbf{n}, \boldsymbol{\mu_k}}\hat{\psi}_{\mathbf{n}+\boldsymbol{\mu}_k,\beta}\frac{1}{2}\varepsilon_{\gamma\delta}\hat{\psi}^\dagger_{\mathbf{n}+\boldsymbol{\mu}_k,\gamma}\hat{\psi}^\dagger_{\mathbf{n}+\boldsymbol{\mu}_k,\delta}\ket{0,0,0}\ket{0}_\mathbf{n}\ket{0}_{\mathbf{n}+\boldsymbol{\mu}_k} =\nonumber \\
       & =\frac{1}{2}\varepsilon_{\gamma\delta}\hat{\psi}_{\mathbf{n},\alpha}^\dagger\hat{U}^{\alpha\beta}_{\mathbf{n}, \boldsymbol{\mu_k}}\bigg(\delta_{\gamma\beta} - \hat{\psi}^\dagger_{\mathbf{n}+\boldsymbol{\mu}_k,\gamma}\hat{\psi}_{\mathbf{n}+\boldsymbol{\mu}_k,\beta}\bigg)\hat{\psi}^\dagger_{\mathbf{n}+\boldsymbol{\mu}_k,\delta}\ket{0,0,0}\ket{0}_\mathbf{n}\ket{0}_{\mathbf{n}+\boldsymbol{\mu}_k}=\nonumber \\
       & =\frac{1}{2}\varepsilon_{\gamma\delta}\hat{\psi}_{\mathbf{n},\alpha}^\dagger\bigg(\hat{U}^{\alpha \gamma}_{\mathbf{n}, \boldsymbol{\mu_k}}\hat{\psi}^\dagger_{\mathbf{n}+\boldsymbol{\mu}_k, \delta}-\hat{U}^{\alpha \delta}_{\mathbf{n}, \boldsymbol{\mu_k}}\hat{\psi}^\dagger_{\mathbf{n}+\boldsymbol{\mu}_k, \gamma}\bigg)\ket{0,0,0}\ket{0}_\mathbf{n}\ket{0}_{\mathbf{n}+\boldsymbol{\mu}_k}=\nonumber                      \\
     & =\frac{1}{2}\varepsilon_{\gamma\delta}\bigg(\ket{\frac{1}{2},\alpha,\gamma}\ket{\alpha}_\mathbf{n}\ket{\delta}_{\mathbf{n}+\boldsymbol{\mu}_k}-\ket{\frac{1}{2},\alpha,\delta}\ket{\alpha}_\mathbf{n}\ket{\gamma}_{\mathbf{n}+\boldsymbol{\mu}_k}\bigg)=\varepsilon_{\gamma\delta}\ket{\frac{1}{2},\alpha,\gamma}\ket{\alpha}_\mathbf{n}\ket{\delta}_{\mathbf{n}+\boldsymbol{\mu}_k},
\end{align}
and similarly
\begin{equation}
	\hat{A}^\dagger\ket{0,0,0}\ket{2}_\mathbf{n}\ket{0}_{\mathbf{n}+\boldsymbol{\mu}_k}=\varepsilon_{\gamma\delta}\ket{\frac{1}{2},\alpha,\gamma}\ket{\delta}_\mathbf{n}\ket{\alpha}_{\mathbf{n}+\boldsymbol{\mu}_k},
\end{equation}
where
\begin{equation}
	\varepsilon_{\gamma\delta}\varepsilon_{\sigma\rho}\braket{\frac{1}{2},\beta,\sigma }{\frac{1}{2},\alpha,\gamma}\braket{\beta}{ \delta}_{\mathbf{n}+\boldsymbol{\mu}_k}\braket{\rho}{\alpha}_\mathbf{n}  = 2,
\end{equation}
and
\begin{equation}
	\varepsilon_{\gamma \delta}\varepsilon_{\sigma \rho}\braket{\frac{1}{2},\beta,\sigma }{ \frac{1}{2},\alpha,\gamma}\braket{\beta }{\alpha}_{\mathbf{n}+\boldsymbol{\mu}_k}\braket{\rho}{\delta}_\mathbf{n} = \varepsilon_{\gamma \delta}\varepsilon_{\sigma\rho}\braket{\frac{1}{2},\beta,\sigma }{\frac{1}{2},\alpha,\gamma}\braket{\beta}{\alpha}_\mathbf{n}\braket{\rho}{\delta}_{\mathbf{n}+\boldsymbol{\mu}_k} = 4,
\end{equation}
where $\braket{\rho}{\alpha}_{\mathbf n}=\delta_{\rho\alpha}$ is the single-fermion color overlap and $\braket{\tfrac12,\beta,\sigma}{\tfrac12,\alpha,\gamma}=\delta_{\beta\alpha}\delta_{\sigma\gamma}$ is the single-link overlap. 
This means that as a single component of $\hat{A}$ acts on the system, it will have $\langle \hat{H}_0 \rangle=\frac{3g^2}{8}$, as $\hat{E}^2\ket{\frac{1}{2}, m_L, m_R}=\frac{1}{2}\cdot\frac{3}{2}\ket{\frac{1}{2}, m_L, m_R}=\frac{3}{4}\ket{\frac{1}{2}, m_L, m_R}$, which means
\begin{align}
    \hat{H}_\text{eff}^{(2)}= & \frac{8}{3g^2}\frac{t^2}{4}\ket{0,0,0}\bra{0,0,0}\Bigg(-\varepsilon_{\gamma\delta}\ket{0}_\mathbf{n}\ket{2}_{\mathbf{n}+\boldsymbol{\mu}_k}\bra{\frac{1}{2},\alpha,\gamma}\bra{\alpha}_\mathbf{n}\bra{\delta}_{\mathbf{n}+\boldsymbol{\mu}_k} + \ket{2}_\mathbf{n}\ket{0}_{\mathbf{n}+\boldsymbol{\mu}_k}\varepsilon_{\gamma\delta}\bra{\frac{1}{2},\alpha,\gamma}\bra{\delta}_\mathbf{n}\bra{\alpha}_{\mathbf{n}+\boldsymbol{\mu}_k} \Bigg)\nonumber \\
    & \Bigg(\varepsilon_{\gamma\delta}\ket{\frac{1}{2},\alpha,\gamma}\ket{\alpha}_\mathbf{n}\ket{\delta}_{\mathbf{n}+\boldsymbol{\mu}_k}\bra{0}_\mathbf{n}\bra{2}_{\mathbf{n}+\boldsymbol{\mu}_k} - \varepsilon_{\gamma\delta}\ket{\frac{1}{2},\alpha,\gamma}\ket{\delta}_\mathbf{n}\ket{\alpha}_{\mathbf{n}+\boldsymbol{\mu}_k} \bra{2}_\mathbf{n}\bra{0}_{\mathbf{n}+\boldsymbol{\mu}_k}\Bigg)=  \\
    = & \frac{2t^2}{3g^2}\Bigg(-4\bigg(\ket{0}_\mathbf{n}\ket{2} _{\mathbf{n}+\boldsymbol{\mu}_k}\bra{0}_\mathbf{n}\bra{2}_{\mathbf{n}+\boldsymbol{\mu}_k} +\ket{2}_\mathbf{n}\ket{0} _{\mathbf{n}+\boldsymbol{\mu}_k}\bra{2}_\mathbf{n}\bra{0}_{\mathbf{n}+\boldsymbol{\mu}_k} \bigg)+\nonumber                 \\ &+2\bigg(\ket{0}_\mathbf{n}\ket{2} _{\mathbf{n}+\boldsymbol{\mu}_k}\bra{2}_\mathbf{n}\bra{0}_{\mathbf{n}+\boldsymbol{\mu}_k} +\ket{2}_\mathbf{n}\ket{0} _{\mathbf{n}+\boldsymbol{\mu}_k}\bra{0}_\mathbf{n}\bra{2}_{\mathbf{n}+\boldsymbol{\mu}_k} \bigg)\Bigg)
\end{align}
\twocolumngrid
Mapping to the spin representation $\ket{0}\equiv\ket{\downarrow}$, $\ket{2}\equiv\ket{\uparrow}$, this can be written as
\begin{equation}
	\hat{H}_\text{eff}^{(2)} = \frac{2t^2}{3g^2}\left(\hat{\boldsymbol{\sigma}}_\mathbf{n}\cdot\hat{\boldsymbol{\sigma}}_{\mathbf{n}+\boldsymbol{\mu}_k} +\hat{\sigma}_\mathbf{n}^z\hat{\sigma}_{\mathbf{n}+\boldsymbol{\mu}_k}^z - 2\right),
	\label{eq:Heff2_XXZ}
\end{equation}
where $\hat{\boldsymbol{\sigma}}_\mathbf{n}\cdot\hat{\boldsymbol{\sigma}}_{\mathbf{n}+\boldsymbol{\mu}_k}=\hat{\sigma}_\mathbf{n}^x\hat{\sigma}_{\mathbf{n}+\boldsymbol{\mu}_k}^x+\hat{\sigma}_\mathbf{n}^y\hat{\sigma}_{\mathbf{n}+\boldsymbol{\mu}_k}^y+\hat{\sigma}_\mathbf{n}^z\hat{\sigma}_{\mathbf{n}+\boldsymbol{\mu}_k}^z$.  
This is an antiferromagnetic XXZ interaction with exchange coupling $J=2t^2/(3g^2)$ and Ising anisotropy $\Delta=2$.
The anisotropy is a direct consequence of Pauli blocking: on a link whose two sites are both empty or both doubly occupied no hopping is possible, since $\hat{A}\ket{2}_\mathbf{n}=\hat{A}^\dagger\ket{0}_\mathbf{n}=0$, and Eq.~\eqref{eq:Heff2_XXZ} indeed vanishes on $\ket{\uparrow\uparrow}$ and $\ket{\downarrow\downarrow}$.
On the anti-aligned links, where $\hat{\sigma}^z_\mathbf{n}\hat{\sigma}^z_{\mathbf{n}+\boldsymbol{\mu}_k}=-\mathbb{I}$, it reduces to the isotropic antiferromagnetic Heisenberg form $\frac{2t^2}{3g^2}\big(\hat{\boldsymbol{\sigma}}_\mathbf{n}\cdot\hat{\boldsymbol{\sigma}}_{\mathbf{n}+\boldsymbol{\mu}_k}-3\big)$ expected from the standard strong-coupling expansion of SU$(2)$ lattice gauge theory with staggered fermions~\cite{krasnitzPhaseStructureLattice1988}.
The ground-state energies $e_{\mathrm{PT}}$ quoted in Table~\ref{tab:high_coupling_benchmark_tab} were obtained by exact diagonalization of the full $\hat{H}_\text{eff}$.

\bibliographystyle{apsrev4-2}
\bibliography{biblio}
\end{document}